\documentclass[twocolumn,twocolappendix]{aastex631}

\usepackage{threeparttable}

\submitjournal{ApJL}

\shorttitle{Duration of early star formation in NGC\,628 with JWST}
\shortauthors{Kim et al.}

\graphicspath{{./}{figures/}}

\begin{document}
\suppressAffiliations
\title{PHANGS-JWST First Results: Duration of the early phase of massive star formation in NGC\,628}

\correspondingauthor{Jaeyeon Kim}
\email{kim@uni-heidelberg.de}

\author[0000-0002-0432-6847]{Jaeyeon Kim}
\affiliation{Zentrum f\"{u}r Astronomie der Universit\"{a}t Heidelberg, Institut f\"{u}r Theoretische Astrophysik, Albert-Ueberle-Str. 2, 69120 Heidelberg}

\author[0000-0002-5635-5180]{M\'elanie Chevance}
\affiliation{Zentrum f\"{u}r Astronomie der Universit\"{a}t Heidelberg, Institut f\"{u}r Theoretische Astrophysik, Albert-Ueberle-Str. 2, 69120 Heidelberg}
\affiliation{Cosmic Origins Of Life (COOL) Research DAO, coolresearch.io}

\author[0000-0002-8804-0212]{J.~M.~Diederik~Kruijssen}
\affiliation{Cosmic Origins Of Life (COOL) Research DAO, coolresearch.io}

\author[0000-0003-0410-4504]{Ashley.~T.~Barnes}
\affiliation{Argelander-Institut f\"{u}r Astronomie, Universit\"{a}t Bonn, Auf dem H\"{u}gel 71, 53121, Bonn, Germany}

\author[0000-0003-0166-9745]{Frank Bigiel}
\affiliation{Argelander-Institut f\"{u}r Astronomie, Universit\"{a}t Bonn, Auf dem H\"{u}gel 71, 53121, Bonn, Germany}

\author[0000-0003-4218-3944]{Guillermo A.~Blanc}
\affiliation{The Observatories of the Carnegie Institution for Science, 813 Santa Barbara St., Pasadena, CA, USA}
\affiliation{Departamento de Astronom\'{i}a, Universidad de Chile, Camino del Observatorio 1515, Las Condes, Santiago, Chile}

\author[0000-0003-0946-6176]{Médéric~Boquien}
\affiliation{Centro de Astronomía (CITEVA), Universidad de Antofagasta, Avenida Angamos 601, Antofagasta, Chile}

\author[0000-0001-5301-1326]{Yixian Cao}
\affiliation{Max-Planck-Institut f\"ur Extraterrestrische Physik (MPE), Giessenbachstr. 1, D-85748 Garching, Germany}

\author[0000-0002-8549-4083]{Enrico Congiu}
\affiliation{Departamento de Astronom\'{i}a, Universidad de Chile, Camino del Observatorio 1515, Las Condes, Santiago, Chile}

\author[0000-0002-5782-9093]{Daniel~A.~Dale}
\affiliation{Department of Physics and Astronomy, University of Wyoming, Laramie, WY 82071, USA}

\author[0000-0002-4755-118X]{Oleg V.~Egorov}
\affiliation{Astronomisches Rechen-Institut, Zentrum f\"{u}r Astronomie der Universit\"{a}t Heidelberg, M\"{o}nchhofstra\ss e 12-14, D-69120 Heidelberg, Germany}

\author[0000-0001-5310-467X]{Christopher M. Faesi}
\affiliation{University of Connecticut, Department of Physics, 196A  Auditorium Road, Unit 3046, Storrs, CT, 06269}

\author[0000-0001-6708-1317]{Simon~C.~O.~Glover}
\affiliation{Zentrum f\"{u}r Astronomie der Universit\"{a}t Heidelberg, Institut f\"{u}r Theoretische Astrophysik, Albert-Ueberle-Str. 2, 69120 Heidelberg}

\author[0000-0002-3247-5321]{Kathryn~Grasha}
\affiliation{Research School of Astronomy and Astrophysics, Australian National University, Canberra, ACT 2611, Australia}   
\affiliation{ARC Centre of Excellence for All Sky Astrophysics in 3 Dimensions (ASTRO 3D), Australia}

\author[0000-0002-9768-0246]{Brent Groves}
\affiliation{International Centre for Radio Astronomy Research, University of Western Australia, 7 Fairway, Crawley, 6009 WA, Australia}

\author[0000-0002-8806-6308]{Hamid Hassani}
\affiliation{Department of Physics, University of Alberta, Edmonton, AB T6G 2E1, Canada}

\author[0000-0002-9181-1161]{Annie~Hughes}
\affiliation{IRAP, Universit\'e de Toulouse, CNRS, CNES, UPS, (Toulouse), France}

\author[0000-0002-0560-3172]{Ralf S.\ Klessen}
\affiliation{Zentrum f\"{u}r Astronomie der Universit\"{a}t Heidelberg, Institut f\"{u}r Theoretische Astrophysik, Albert-Ueberle-Str. 2, 69120 Heidelberg}
\affiliation{Universit\"{a}t Heidelberg, Interdisziplin\"{a}res Zentrum f\"{u}r Wissenschaftliches Rechnen, Im Neuenheimer Feld 205, D-69120 Heidelberg, Germany}

\author[0000-0001-6551-3091]{Kathryn Kreckel}
\affiliation{Astronomisches Rechen-Institut, Zentrum f\"{u}r Astronomie der Universit\"{a}t Heidelberg, M\"{o}nchhofstra\ss e 12-14, D-69120 Heidelberg, Germany}

\author[0000-0003-3917-6460]{Kirsten L. Larson}
\affiliation{AURA for the European Space Agency (ESA), Space Telescope Science Institute, 3700 San Martin Drive, Baltimore, MD 21218, USA}

\author[0000-0003-0946-6176]{Janice C. Lee}
\affiliation{Gemini Observatory/NSF's NOIRLab, 950 N. Cherry Avenue, Tucson, AZ, USA}
\affiliation{Steward Observatory, University of Arizona, 933 N Cherry Ave,Tucson, AZ 85721, USA}

\author[0000-0002-2545-1700]{Adam~K.~Leroy}
\affiliation{Department of Astronomy, The Ohio State University, 140 West 18th Avenue, Columbus, Ohio 43210, USA}
\affiliation{Center for Cosmology and Astroparticle Physics, 191 West Woodruff Avenue, Columbus, OH 43210, USA}

\author[0000-0001-9773-7479]{Daizhong Liu}
\affiliation{Max-Planck-Institut f\"ur Extraterrestrische Physik (MPE), Giessenbachstr. 1, D-85748 Garching, Germany}

\author[0000-0001-6353-0170]{Steven~N.~Longmore}
\affiliation{Astrophysics Research Institute, Liverpool John Moores University, IC2, Liverpool Science Park, 146 Brownlow Hill, Liverpool L3 5RF, UK}
\affiliation{Cosmic Origins Of Life (COOL) Research DAO, coolresearch.io}

\author[0000-0002-6118-4048]{Sharon E. Meidt}
\affiliation{Sterrenkundig Observatorium, Universiteit Gent, Krijgslaan 281 S9, B-9000 Gent, Belgium}

\author[0000-0002-1370-6964]{Hsi-An Pan}
\affiliation{Department of Physics, Tamkang University, No.151, Yingzhuan Road, Tamsui District, New Taipei City 251301, Taiwan}

\author[0000-0003-3061-6546]{Jérôme Pety}
\affiliation{IRAM, 300 rue de la Piscine, 38400 Saint Martin d'H\`eres, France}
\affiliation{LERMA, Observatoire de Paris, PSL Research University, CNRS, Sorbonne Universit\'es, 75014 Paris}

\author[0000-0002-0472-1011]{Miguel~Querejeta}
\affiliation{Observatorio Astron\'{o}mico Nacional (IGN), C/Alfonso XII, 3, E-28014 Madrid, Spain}

\author[0000-0002-5204-2259]{Erik~Rosolowsky}
\affiliation{Department of Physics, University of Alberta, Edmonton, AB T6G 2E1, Canada}

\author[0000-0002-2501-9328]{Toshiki Saito}
\affiliation{National Astronomical Observatory of Japan, 2-21-1 Osawa, Mitaka, Tokyo, 181-8588, Japan}

\author[0000-0002-4378-8534]{Karin Sandstrom}
\affiliation{Center for Astrophysics and Space Sciences, Department of Physics, University of California, San Diego\\9500 Gilman Drive, La Jolla, CA 92093, USA}

\author[0000-0002-3933-7677]{Eva Schinnerer}
\affiliation{Max-Planck-Institut f\"{u}r Astronomie, K\"{o}nigstuhl 17, D-69117, Heidelberg, Germany}

\author[0000-0002-0820-1814]{Rowan~J.~Smith}
\affiliation{Jodrell Bank Centre for Astrophysics, Department of Physics and Astronomy, University of Manchester, Oxford Road, Manchester M13 9PL, UK}

\author[0000-0003-1242-505X]{Antonio~Usero}
\affiliation{Observatorio Astron\'{o}mico Nacional (IGN), C/Alfonso XII, 3, E-28014 Madrid, Spain}

\author[0000-0002-7365-5791]{Elizabeth J. Watkins}
\affiliation{Astronomisches Rechen-Institut, Zentrum f\"{u}r Astronomie der Universit\"{a}t Heidelberg, M\"{o}nchhofstra\ss e 12-14, D-69120 Heidelberg, Germany}

\author[0000-0002-0786-7307]{Thomas G. Williams}
\affiliation{Sub-department of Astrophysics, Department of Physics, University of Oxford, Keble Road, Oxford OX1 3RH, UK}
\affiliation{Max-Planck-Institut f\"{u}r Astronomie, K\"{o}nigstuhl 17, D-69117, Heidelberg, Germany}

\begin{abstract}
The earliest stages of star formation, when young stars are still deeply embedded in their natal clouds, represent a critical phase in the matter cycle between gas clouds and young stellar regions. Until now, the high-resolution infrared observations required for characterizing this heavily obscured phase (during which massive stars have formed, but optical emission is not detected) could only be obtained for a handful of the most nearby galaxies. One of the main hurdles has been the limited angular resolution of the \textit{Spitzer Space Telescope}. With the revolutionary capabilities of the JWST, it is now possible to investigate the matter cycle during the earliest phases of star formation as a function of the galactic environment. In this Letter, we demonstrate this by measuring the duration of the embedded phase of star formation and the implied time over which molecular clouds remain inert in the galaxy NGC\,628 at a distance of 9.8~Mpc, demonstrating that the cosmic volume where this measurement can be made has increased by a factor of $>100$ compared to \textit{Spitzer}. We show that young massive stars remain embedded for $5.1_{-1.4}^{+2.7}$\,Myr ($2.3_{-1.4}^{+2.7}$\,Myr of which being heavily obscured), representing $\sim 20 \% $ of the total cloud lifetime. These values are in broad agreement with previous measurements in five nearby ($D < 3.5$\,Mpc) galaxies and constitute a proof of concept for the systematic characterization of the early phase of star formation across the nearby galaxy population with the PHANGS--JWST survey.
\end{abstract}

\keywords{Galaxies (573) --- Star formation (1569) --- Molecular clouds (1072) --- Interstellar medium (847)}

\section{Introduction}
Over the last two decades, a growing number of multi-wavelength, cloud-scale observations have revealed a spatial offset between cold molecular gas and H\,{\sc ii} regions in galaxies \citep{engargiola03, blitz07, kawamura09, onodera10, schruba10, miura12, meidt15, corbelli17, kruijssen19, schinnerer19, barnes20, pan22}. The statistical characterization of this offset has enabled a quantitative description of the evolutionary lifecycle of giant molecular clouds (GMCs), during which gas is turning into stars \citep{kruijssen19, chevance20_rev, chevance20, zabel20, kim21, kim22, chevance22, chevance22_rev, lu22, ward22}. These studies have illustrated that GMCs are transient objects that survive for $1{-}3$ dynamical timescales ($10{-}30$\,Myr, with typical associated uncertainties of $\sim$25\%) and are dispersed quickly by feedback from newly formed stars, after a long phase during which GMCs appear inert and devoid of massive stars ($70{-}90$\% of the cloud lifetime), before the star formation is detected through H$\alpha$ emission. In these studies, GMCs have masses over $10^{4}{-}10^5\,M_{\odot}$ and the lifetimes of these objects represent the time they spend being bright in CO emission, until the molecular gas has been dispersed by the resulting H{\sc ii} region. 

However, the earliest phases of star formation are heavily embedded and invisible in H$\alpha$ due to the extinction from the surrounding dense gas and dust. Therefore, the duration of these phases and the time that clouds spend being truly inert are still poorly constrained, and so is the time needed for the feedback from these heavily embedded stars to blow out enough of the natal cloud to enable the detection of H$\alpha$ emission. This limits our understanding of the physical mechanisms playing a role in the first stages of star formation. Measuring these characteristic timescales is crucial to establish which mechanisms are responsible for dispersing the molecular clouds \citep[e.g.][]{lopez14} and for distinguishing whether star formation is delayed by the decay of initial turbulence \citep[e.g.][]{klessen16, padoan17} or suppressed by galactic-scale dynamics, such as the shear associated with spiral arms and differential rotation preventing collapse of the clouds \citep[e.g.][]{meidt18}. 

High resolution infrared observations ($\sim$1\,pc scale) of star-forming regions in the Milky Way have revealed that molecular clouds spend $30{-}40$\% of their lifetime with embedded stars \citep{lada03, battersby17}. Massive proto-clusters ($\sim\,10^{4}\,M_{\odot}$) in our galaxy are actively forming stars and appear to have a very short starless phase ($<0.5$\,Myr; \citealp{ginsburg12}). In nearby galaxies, the timescales between successive stages of the gas-to-stars evolutionary cycle can be estimated by combining ages of star clusters with distances between these clusters and their neighboring GMCs. These results suggest that the embedded star-forming phase lasts for $2{-}5$\,Myr, of which the initial $0{-}2$\,Myr are heavily obscured, i.e.\ on-going star formation is detected in mid-infrared or in radio continuum but invisible in H$\alpha$ and ultraviolet emission \citep{whitmore14, corbelli17, calzetti15, grasha18, turner22}. 

\citet{kruijssen14} and \citet{kruijssen18} have introduced a statistically rigorous method that translates the observed spatial decorrelation between cold gas and star formation rate (SFR) tracers into their underlying timescales. In \citet{kim21}, this method has been applied to six nearby star-forming galaxies using CO, \textit{Spitzer} 24\,$\mu$m, and H$\alpha$ emission maps, tracing molecular clouds, embedded star formation, and exposed star formation, respectively. This provided systematic constraints on the duration of the embedded phase of star formation for five of these six galaxies, which was shown to last for $2{-}7$\,Myr, constituting $20{}-50$\% of the cloud lifetime. The first half of this phase is heavily obscured and only detected in CO and 24\,$\mu$m, while being invisible in H$\alpha$ emission. Until now, the number of galaxies where we could constrain these timescales was restricted to these five galaxies, with distances of $D < 3.5$\,Mpc. This small sample was due to the limited resolution of the \textit{Spitzer} 24\,$\rm\mu m$ observations (6\arcsec) and the requirement that the observations need to resolve each galaxy into its distinctive units of star formation (e.g.\ GMCs and H~\textsc{ii} regions, typically separated by $\sim100$\,pc). The results of deconvolution algorithms \citep{backus05} applied to more distant galaxies (M51 at 8.6\,Mpc; \citealt{dumas11}) did not lead to a sufficient data quality to successfully perform this measurement.

The Mid-Infrared Instrument (MIRI) aboard JWST has opened a new era of infrared astronomy with unprecedented spatial resolution and sensitivity in the mid-infrared. In particular, observations at 21\,$\mu$m tracing embedded young stellar populations reach a resolution of 0.67\arcsec,\ allowing the cloud-to-star life cycle to be characterized, with the above method, out to considerably larger distances of up to 25 Mpc. The PHANGS\footnote{The Physics at High Angular resolution in Nearby GalaxieS project: http://phangs.org} collaboration is carrying out the PHANGS--JWST survey (\citealp{LEE_PHANGSJWST} this Issue; Program ID 02107) to map the star-forming disk of 19 galaxies in a wide range of wavelengths, from 2\,$\mu$m to 21\,$\mu$m. This translates into a physical scale of $20-60$\,pc in the 21\,$\mu$m band for the galaxies in this sample (at distances between 5\,Mpc and 20\,Mpc). So far, four of these galaxies have been observed (IC\,5332, NGC\,628, NGC\,1365, NGC\,7496) with MIRI JWST. In this Letter, we extend our previous analysis by \citet{kim21} by characterizing the duration of the early phase of star formation in one of these initial galaxies, NGC\,628, which is the most nearby (yet 3 times further away than the most distant galaxy analyzed in \citealp{kim21}), and for which the duration of the CO- and H$\alpha$-bright phases have already been obtained in our previous works \citep{chevance20, kim22}. Following previous works using \textit{Spitzer} 24\,$\mu$m as a tracer for embedded massive stars \citep{corbelli17, calzetti15, kim21}, we define the duration of `embedded star formation' probed at 21\,$\mu$m with JWST as the total phase during which CO and 21\,$\mu$m are found to be overlapping, whereas the `heavily obscured phase' refers to the phase where both CO and 21\,$\mu$m are detected \textit{without} associated H$\alpha$ emission.

\section{Observations}\label{sec:obs}

In order to trace embedded star formation, we use the 21\,$\mu$m emission map observed with MIRI on board JWST as a part of PHANGS--JWST survey. This data was obtained from the Mikulski Archive for Space Telescopes (MAST) at the Space Telescope Science Institute\footnote{The specific observations analyzed can be accessed via \dataset[DOI]{http://dx.doi.org/10.17909/9bdf-jn24}.}. This mid-infrared wavelength has been widely used as a tracer of embedded star formation, because a substantial fraction of the emission, especially that with compact morphology, originates from dust excitation by radiation from surrounding massive stars and empirically exhibits a correlation with tracers of massive star formation (\citealp{kennicutt12, galliano18,HASSANI_PHANGSJWST, LEROY1_PHANGSJWST, THILKER_PHANGSJWST}). In particular, using four initial targets,  \citet{HASSANI_PHANGSJWST} have found that 90\% of compact 21\,$\mu$m sources are associated with H{\sc ii} regions detected in extinction corrected H$\alpha$ maps from MUSE.  Furthermore, \citet{HASSANI_PHANGSJWST} have shown that background galaxies and evolved stars identified in the 21\,$\mu$m map are faint, only constituting $\sim$3\% of the total 21\,$\mu$m emission flux and therefore they are unlikely to affect our measurements, because the quantities constrained with our methodology are flux-weighted (see Section~\ref{sec:method}). The JWST map has a physical resolution of $\sim 30$\,pc at the distance of NGC\,628 (9.84\,Mpc; \citealp{anand21, anand21_2}) and a 1$\sigma$ surface brightness sensitivity of $\sim$0.3\,$\rm MJy\,sr^{-1}$ at the native resolution of 0.67\arcsec. Details on the data reduction can be found in \citet{LEE_PHANGSJWST}.

\begin{figure*}
\includegraphics[scale=0.87]{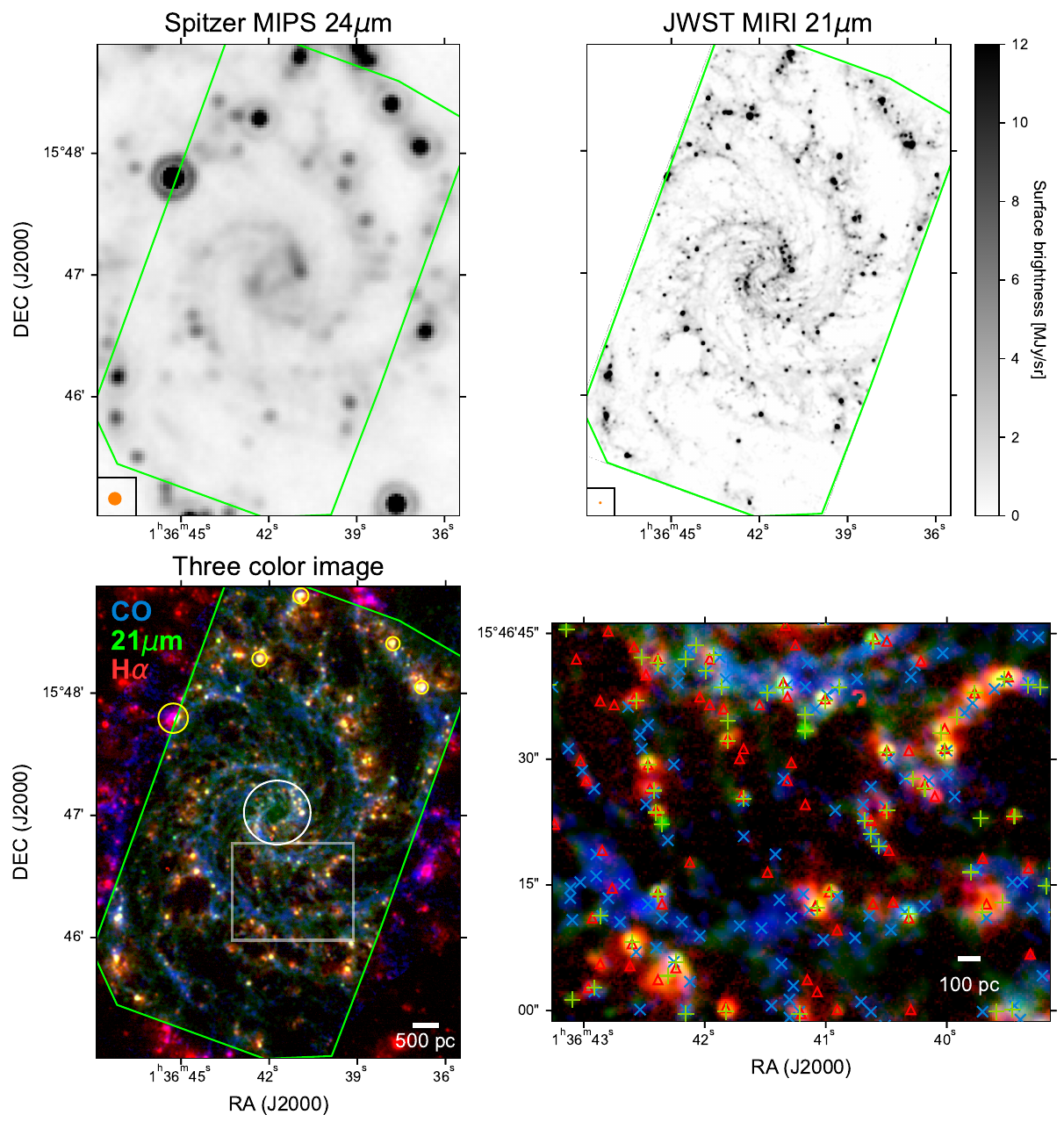}
\caption{\textit{Top:} Comparison between the \textit{Spitzer} 24\,$\mu$m map (left) and the JWST 21\,$\mu$m map (middle), which has 10 times better resolution (0.67\arcsec) compared to \textit{Spitzer} (6\arcsec). Orange circles show the beam in each panel. \textit{Bottom:} Composite three color images obtained by combining CO (blue), 21\,$\mu$m (green), and H$\alpha$ (red). The bottom right panel shows the zoomed-in image of the white rectangular region marked in the bottom left panel, with symbols indicating the distribution of emission peaks using the same color scheme. Emission peaks of CO (indicated with $\textcolor{blue}{\times}$), 21\,$\mu$m ($\textcolor{green}{+}$), and H$\alpha$ ($\textcolor{red}{\triangle}$) show spatial offsets, indicating that these represent distinctive stages of star formation. The JWST field of view is outlined in green. The crowded galactic center (white circle), as well as extremely bright star-forming regions (yellow circles) are excluded from our analysis (see text).}\label{fig:map}
\end{figure*}

In Figure~\ref{fig:map}, we show a comparison between the \textit{Spitzer} MIPS map at 24\,$\mu$m and the JWST MIRI map at 21\,$\mu$m of NGC\,628. The increase in resolution by a factor of almost 10 allows us to resolve individual regions in the galaxy. A composite three-color image of the CO, 21\,$\mu$m, and H$\alpha$ emission maps is also provided, where the spatial small-scale decorrelation of these tracers is illustrated by the color variations.The H$\alpha$ emission map is from PHANGS--H$\alpha$ (Preliminary version; A.~Razza et al.\ in prep.) observed using the Wide Field Imager instrument at the MPG-ESO 2.2-m telescope at the La Silla Observatory.

We use the $^{12}$CO($J{=}2{-}1$) transition (CO hereafter) from PHANGS--ALMA as a tracer of molecular gas. A detailed description of the full sample and data reduction can be found in \citet{leroy21_survey} and \citet{leroy21_pipe}. The observations were carried out with the 12-m array, as well as with the 7-m and total power antennas of the Atacama Large Millimeter/submillimeter Array (ALMA). We use the moment-0 map at the native resolution reduced with an inclusive signal masking scheme with high completeness (the ``broad'' mask; see \citealp{leroy21_pipe}) The resulting CO map has a resolution of 1.12\arcsec ($\sim50$\,pc) and a 5$\sigma$ molecular gas mass sensitivity of $5\times10^{4}\,M_{\odot}$ \citep{leroy21_survey}. After the removal of diffuse emission  (see Section~\ref{sec:method}), the faintest identified CO emission peak has a mass of $10^{5}\,M_{\odot}$.
 
In order to perform the next steps of the analysis (see Section~\ref{sec:method}), we first convolve and then reproject the 21\,$\mu$m emission map to match the resolution and the pixel grid of the CO map. During the convolution, we use a kernel that translates the JWST MIRI point spread function to a Gaussian, matched to the beam of the CO map and generated using the method of \citet{aniano11}. Our statistical method (described in Section~\ref{sec:method}) makes use of the relative spatial distribution of the molecular clouds and young stellar regions to derive their associated time-scales and therefore the astrometric precision of the CO and 21\,$\mu$m map must be sufficient to detect offsets. Extensive experiments of the method with simulated data show that an acceptable astrometric precision is $1/3$ of the beam \citep{kruijssen18, hygate19}, which corresponds here to $\sim 0.4$\arcsec.\ \citet{LEE_PHANGSJWST} have shown that MIRI images, aligned using asymptotic giant branch stars and PHANGS--HST data \citep{lee22}, have a astrometric uncertainties of $\pm 0.1$\arcsec,\ comfortably satisfying the required precision. The astrometric precision of the H$\alpha$ map is also measured to be within the acceptable precision with $0.1\arcsec{-}0.2\arcsec$ by matching stellar sources to Gaia DR2 catalog \citep{gaiadr2} or SINGS and WFI data \citep[][Razza et al. in prep.]{chevance20}. 

Following our previous analysis (e.g.\ \citealp{kim21, kim22}), we further mask very bright regions that can potentially bias our measurements of timescales (yellow circles in Figure~\ref{fig:map}). These bright peaks represent outliers in the luminosity function of the peaks identified using \textsc{Clumpfind} \citep{williams94} and also seen in \citet{HASSANI_PHANGSJWST}. The galactic center (white circle) is also excluded from our analysis, because crowding of sources makes it difficult to identify star-forming regions and molecular clouds in this environment.

\section{Method}\label{sec:method}
We now briefly describe our analysis method (the `uncertainty principle for star formation', formalized in the \textsc{Heisenberg}\footnote{The \textsc{Heisenberg} code is publicly available at \url{https://github.com/mustang-project/Heisenberg}.} code) and the main input parameters used. We refer readers to \citet{kruijssen18} for a full description and rigorous validation of the code using simulated galaxies and \citet{kruijssen14} for an introduction of the method. This method has been applied to $\sim$60 observed galaxies \citep{kruijssen19b, chevance20, chevance22, haydon20, zabel20, ward20_HI, lu22, ward22, kim21, kim22}, including NGC\,628, using CO and H$\alpha$ as tracers of molecular gas and SFR. Unless stated otherwise, here we adopt the same input parameters for this galaxy as in \citet{chevance20} and \citet{kim22}, describing the main properties of the galaxy and the CO and H$\alpha$ observations. 

Our method exploits the relative spatial distributions of tracers of successive phases of the evolution from GMCs to young stellar regions. Contrary to the observed tight correlation on $\sim$kpc scales between molecular gas and SFR tracers (e.g.\ CO and H$\alpha$) that defines the well known ``star formation relation'' (e.g.\ \citealp{kennicutt98, bigiel08}), small-scale ($\sim$100\,pc) observations resolving galaxies into independent star-forming regions and clouds reveal spatial offsets between them, increasing the observed scatter of the star formation relation. This small-scale decorrelation can be naturally explained by galaxies being composed of `independent' regions, each undergoing independent evolution from molecular cloud assembly to star formation and feedback, which disperses the natal clouds and leaves young stellar regions without associated molecular gas \citep{schruba10,onodera10,kruijssen14}. Our methodology assumes that the spatial distribution of such regions is locally isotropic on the scale of the mean separation length between regions (a few $100$\,pc), which is the largest scale that our measurements are sensitive to. This means that our measurements are not affected by galactic-morphological features, such as gaseous spiral arms that produce linear features on kpc scales \citep{kruijssen18}.

To translate the observed decorrelation of cold gas and star formation tracers into the underlying evolutionary timescales associated with each tracer \citep{kruijssen18}, we first identify peaks in the CO and 21\,$\mu$m maps using \textsc{Clumpfind} \citep{williams94}. This algorithm uses contours on the map for a set of flux levels separated by a step size $\rm{\delta}log_{10}\mathcal{F}$, with a full range $\rm{\Delta}log_{10}\mathcal{F}$ starting from the maximum flux level. For the 21\,$\mu$m emission map, we adopt $\rm{\delta}log_{10}\mathcal{F} = 0.05$ and $\rm{\Delta}log_{10}\mathcal{F}=2.0$, where the choice of this full range is well justified given the distribution of 21\,$\mu$m peaks in \citet[][after excluding bright outliers]{HASSANI_PHANGSJWST}. For the CO emission map, we adopt  $\rm{\delta}log_{10}\mathcal{F} = 0.05$ and $\rm{\Delta}log_{10}\mathcal{F}=1.1$, similar to our previous analysis \citep{chevance20, kim22}. On each identified peak, we then center apertures with a range of sizes from the cloud scale ($l_{\rm ap, min}=50\rm\,pc$, similar to 1\,beamsize) to the galactic scale ($l_{\rm ap, max}=1.5\rm\,kpc$). For each aperture size, we measure the deviation of the gas-to-SFR tracer flux ratio around all peaks compared to the galactic average value.

We then fit an analytical function \citep[see Sect.~3.2.11 of][]{kruijssen18} to the measured flux ratios as a function of aperture size, which depends on the relative duration of emission of each tracer, the relative duration for which they overlap, and the typical separation length between independent regions ($\lambda$). The absolute values of the timescales are obtained by multiplying the best-fitting relative timescale by a known reference timescale ($t_{\rm ref}$). Here, we use the cloud lifetime ($t_{\rm CO}$) derived in our previous analysis \citep{chevance20, kim22} as $t_{\rm ref}$. Since we mask four regions that are extremely bright in 21\,$\mu$m (see Section~\ref{sec:obs} and Figure~\ref{fig:map}), we repeat our previous analysis of NGC\,628 using narrow-band H$\alpha$ as a SFR tracer \citep{chevance20, kim22} to see how the masking impacts the measurements. The results are consistent within the uncertainties, and our new measurements for the masked map are shown in Table~\ref{tab:result} (discussed below). Masking additional bright regions only results in negligible differences in our measurements of the timescales obtained with H$\alpha$ emission (again within uncertainties).
\begin{table}
\begin{center}
\caption{Derived characteristic properties of the evolutionary cycle traced by the CO and H$\alpha$ emission maps, as well as the CO and 21\,$\mu$m emission maps. \label{tab:result}}
\setlength\tabcolsep{30pt}
\begin{threeparttable}
\begin{tabular}{lc}
\hline
\multicolumn2l{H$\alpha$ as a SFR tracer}\\
\hline
$t_{\rm CO}$ & $23.9_{-2.8}^{+2.5}$~Myr\\
$t_{\rm fb, H\alpha}$ & $2.7_{-0.6}^{+0.5}$~Myr\\
$\lambda_{\rm H\alpha}$ & $96_{-11}^{+13}$~pc\\

\hline
\multicolumn{2}{l}{21\,$\mu$m as a SFR tracer}\\
\hline
$t_{\rm 21\,\mu m}$ & $8.8_{-1.4}^{+3.6}$~Myr\\
$t_{\rm fb, 21\,\mu m}$ & $5.1_{-1.4}^{+2.7}$~Myr\\
$\lambda_{\rm 21\,\mu m}$ & $90_{-17}^{+51}$~pc\\
\hline
\multicolumn{2}{l}{Duration of heavily obscured phase}\\
\hline
$t_{\rm obsc}$ &$2.3_{-1.4}^{+2.7}$\,Myr\\
\hline
\end{tabular}
\end{threeparttable}
\end{center}
\end{table}

Having obtained the lifetime of the CO-bright emission, we can derive the absolute lifetimes of the $21\,\mu$m emission. The fitted model is described by three independent quantities: the timescale over which CO and the SFR tracer are found to be overlapping ($t_{\rm fb, 21\,\mu m}$), the 21$\,\mu$m emitting timescale ($t_{21\,\mu m}$), and the typical separation between independent regions ($\lambda_{\rm 21\mu m}$). The overlapping timescale represents the time over which embedded massive star formation takes place, as well as the time it takes for stellar feedback to disperse the surrounding gas. The fit to the observations returns a three-dimensional probability distribution function (PDF) of the free parameters, which is then marginalized to obtain the one-dimensional PDF of each parameter. The uncertainties quoted here are defined as the $32^{nd}$ percentile of the part of the PDF below the best-fitting value, and the $68^{th}$ percentile of the part of the PDF above the best-fitting value \citep{kruijssen18}.

As part of the analysis process, we filter out potential diffuse emission in both CO and 21\,$\mu$m maps using the method presented in \citet{hygate19}. This is necessary as the presence of diffuse emission can bias our measurements by adding a reservoir of large-scale emission that is not associated with the identified peaks within the aperture, and therefore does not participate in the cycling that is being characterized here. Similarly to our previous analysis of NGC\,628 \citep{chevance20, kim22}, we iteratively remove emission on scales larger than 15$\,\lambda_{\rm 21\mu m}$ using a Gaussian high-pass filter in Fourier space. The threshold of 15$\,\lambda_{\rm 21\mu m}$ was chosen to ensure that the flux loss from the compact region is about 10\%, following the prescription by \citep{hygate19, kruijssen19, chevance20,kim21,kim22}. However, we note that using a higher or lower multiples of $\lambda_{\rm 21\mu m}$ (from 10 to 20) do not significantly impact our measurements (within 1$\sigma$ uncertainties).

In the CO map, filtering extended structures (that constitute $\sim$60\% of the emission) results in lowering the signal-to-noise ratio of small, faint clouds, allowing us to focus on molecular clouds that are likely to form massive stars. Before filtering, the faintest identified CO emission peak has a mass of $10^4\,M_{\odot}$, whereas after filtering the faintest identified CO emission peak has a mass of $ 10^5\,M_{\odot}$, which is likely to give birth to massive stars when assuming a standard initial mass function. In the 21\,$\mu$m map, this removes large-scale emission (constituting $\sim$50\% of the total emission) originating from the interstellar radiation field, which is not related to recent massive star formation but has a non-negligible contribution to the dust heating \citep{draine07, verley09}. \citet{THILKER_PHANGSJWST} report a fraction of mid-infrared emission arising from filamentary structures ($\sim$30\%) that is qualitatively similar to the $\sim$50\% obtained here. The fraction of large-scale emission removed is also broadly consistent with the contribution of the interstellar radiation field to \textit{Spitzer} 24\,$\mu$m wavelength measured in the Milky Way and in Local Group galaxies \citep[$20\%{-} 85\%$][]{koepferl15, viaene17, williams19}. \citet{leroy12} and \citet{LEROY1_PHANGSJWST} also measure $40{-}60$\% of the mid-infrared emission to originate from molecular gas heated by interstellar radiation field. 

\section{Results}

\begin{figure*}
\includegraphics[scale=0.56]{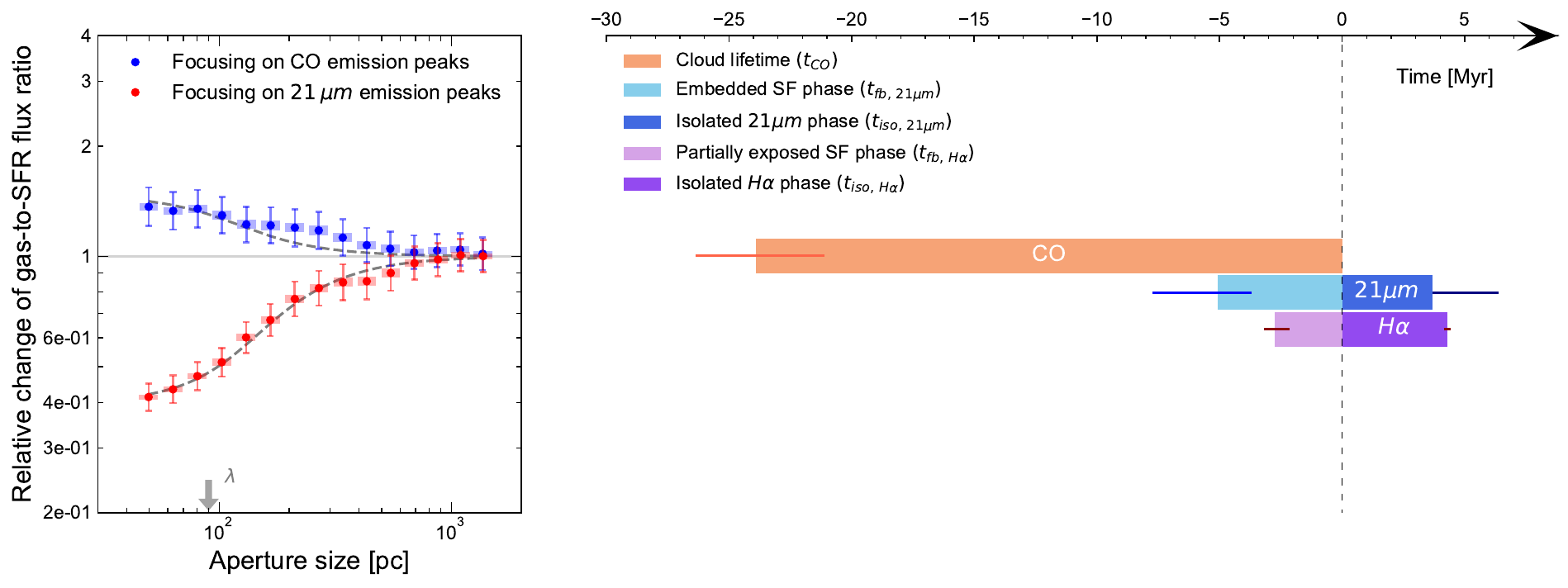}
\caption{The left panel shows the measured deviation of gas-to-SFR tracer (CO-to-21\,$\mu$m) flux ratios compared to the galactic average as a function of the size of apertures centered on CO and 21\,$\mu$m emission peaks. The data underlying this figure can be found in Appendix~\ref{app:data}. The error bars show the 1$\sigma$ uncertainty of each measurement whereas the shaded region within the error bar indicates the effective 1$\sigma$ error, considering the covariance between data points. Our best-fitting model (dashed line), as well as the galactic average (horizontal line) are also shown. The constrained $\lambda$ is indicated with a downward arrow and other best-fitting parameters ($t_{\rm 21\,\mu m}$ and $t_{\rm fb, 21\,\mu m}$) are listed in Table~\ref{tab:result}. The right panel illustrates the evolutionary sequence from inert molecular clouds to embedded star formation, partially exposed star formation, and finally to fully revealed young stellar regions. The duration of the CO emitting phase ($t_{\rm CO}$) is shown in orange while the time during which 21\,$\mu$m and H$\alpha$ emission are detected without associated CO emission are shown in dark blue and dark purple, respectively. The feedback timescale, which is the time for which both CO and a SFR tracer are found coincident, is shown in light blue (for 21\,$\mu$m) and light purple (for H$\alpha$). The error bars indicate the 1$\sigma$ uncertainty for each measurement.} \label{fig:result}
\end{figure*}

Table~\ref{tab:result} lists results from the application of our method to the CO and 21\,$\mu$m maps, tracing molecular gas and embedded star formation, respectively. In Figure~\ref{fig:result} (left) and Appendix~\ref{app:data}, we present the measured variation of gas-to-SFR tracer flux ratio compared to the galactic average, as a function of aperture size. Towards small scales, the flux ratios increasingly diverge from the galactic average, illustrating the spatial decorrelation between CO and 21\,$\mu$m emission on cloud scales. The right panel of Figure~\ref{fig:result} shows the constrained timeline after combining our results for both SFR tracers. At first, clouds are only detected in CO emission for a duration of $t_{\rm CO}-t_{\rm fb, 21\,\mu m}=18.8_{-3.6}^{+2.7}$\,Myr. Then, after the onset of the heavily obscured phase of star formation, 21\,$\mu$m emission is detected together with CO emission (but without associated H$\alpha$) for $t_{\rm obsc}= 2.3_{-1.4}^{+2.7}$\,Myr. Feedback from these newly-formed stars progressively disperses the surrounding gas, revealing young stars emerging from their natal GMC in H$\alpha$ emission for $t_{\rm fb, H\alpha}=2.7_{-0.6}^{+0.5}$\,Myr. Finally, the molecular gas is completely dispersed, leaving only the young stellar regions to be detected through both SFR tracers for about $t_{\rm 21\,\mu m}-t_{\rm fb,21\,\mu m}$ or $t_{\rm H\alpha}-t_{\rm fb,H\alpha}$ of $\sim$4\,Myr on average. In Appendix~\ref{app:acc}, we verify that the measured timescales are reliable with an accuracy of 30\% or better. This implies that these measurements achieve a similar level of confidence as those for the more nearby galaxies presented in \citet{kim21}, despite having been made for a galaxy at a much greater distance.

\subsection{Duration of the embedded ($t_{\rm fb,21\,\mu m}$) and heavily obscured ($t_{\rm obsc}$) phases of star formation}
Because star formation can only continue until molecular clouds have been dispersed, we define the duration of the embedded phase of star formation as the time during which CO and 21\,$\mu m$ emission are found to be overlapping (i.e.\ the feedback timescale, $t_{\rm fb,21\,\mu m}$). We measure $t_{\rm fb,21\,\mu m} = 5.1_{-1.4}^{+2.7}$\,Myr in NGC\,628, which represents $\sim$20\% of the cloud lifetime ($t_{\rm CO}$). These two values fall into the range of those constrained in five nearby galaxies ($2{-}7$\,Myr, $20{-}50$\%) by \citet{kim22}.

The feedback timescale measured with 21\,$\mu$m ($5.1_{-1.4}^{+2.7}$\,Myr) is longer than the one obtained using H$\alpha$ ($2.7_{-0.6}^{+0.5}$\,Myr) as an SFR tracer. This difference can be explained by the fact that the earliest stages of star formation are invisible in H$\alpha$ emission due to the extinction from the surrounding dense gas and dust, while 21\,$\mu$m is detected as it originates from the re-emission of absorbed stellar light by small dust grains \cite[e.g.][]{kennicutt07, galliano18}. We find that this heavily obscured phase of star formation ($t_{\rm obsc}=t_{\rm fb,21\,\mu m}-t_{\rm fb, H\alpha}$) lasts for $2.3_{-1.4}^{+2.7}$\,Myr \footnote{The uncertainties on $t_{\rm obsc}$ are obtained using formal error propagation and therefore are similar to those on $t_{\rm fb,21\,\mu m}$, which shows larger errors than $t_{\rm fb, H\alpha}$. These uncertainties on $t_{\rm obsc}$ should be considered as an upper limit, because no covariance between $t_{\rm fb, H\alpha}$ and $t_{\rm fb,21\,\mu m}$ is assumed, which is unlikely to be true.}, showing a good agreement with the range of values constrained in five nearby galaxies ($1{-}4$\,Myr) by \citet{kim21}. 

The short durations of $t_{\rm fb,21\,\mu m}$ and $t_{\rm obsc}$ support our previous claim that pre-supernova feedback likely drives the dispersal of molecular clouds, as SNe take longer to detonate \citep[$4{-}20$\,Myr;][also see \citealt{barnes22, dellabruna22}]{chevance20, chevance22}. Similar values of $t_{\rm fb,21\,\mu m}$ and $t_{\rm obsc}$ have been measured using ages of stellar clusters and their association with neighboring GMCs \citep{whitmore14, grasha18, grasha19}, as well as using HII region morphologies \citep{hannon19, hannon22}. We note that the ``heavily obscured phase'' is also referred to as ``embedded'' in other works in this Issue, which report qualitatively similar durations of this phase \citep{RODRIGUEZ_PHANGSJWST, WHITMORE_PHANGSJWST}.

\subsection{Duration of the total 21\,$\mu$m emitting phase}
In NGC\,628, we measure the total duration of the 21\,$\mu$m emitting phase ($t_{\rm 21\,\mu m}$) to be $8.8_{-1.4}^{+3.6}$\,Myr, which falls into the range of our previous measurements of this timescale in nearby galaxies \citep[$4{-}14$\,Myr;][]{kim21}. After the star formation is terminated by the dispersal of molecular clouds, the emission at 21\,$\mu$m can still be detected for $\sim$4\,Myr, due to the remaining dust in the HII region, which is heated by the high mass stars that have not yet ended their lives. As shown in Figure~\ref{fig:result}, the end of this isolated 21\,$\mu$m emitting phase (after CO has disappeared) corresponds broadly to the end of the H$\alpha$ emitting phase, indicating that our diffuse emission-filtered 21\,$\mu$m map effectively traces emission related to recent massive star formation. Furthermore, as shown in Figure~\ref{fig:result}, almost 80\% of the total 21\,$\mu$m emitting time-scale coincides with the H$\alpha$ emitting time-scale, showing a good agreement with \citet{HASSANI_PHANGSJWST} who find that 90\% of the 21\,$\mu$m emission peaks in four initial PHANGS--JWST galaxies are associated with H$\alpha$ emission. Our result also agrees with those of \citet{linden22}, who find that 80\% of young massive star cluster candidates identified with JWST near-infrared emission also have an optical counterpart. This also explains why 21\,$\mu$m and H$\alpha$ emission show a tighter correlation than 21\,$\mu$m and CO emission \citep{LEROY1_PHANGSJWST}.

\subsection{Characteristic distance between independent regions}
Figure~\ref{fig:result} shows that GMCs and young stellar regions are spatially decorrelated on small scales, illustrating that galaxies are composed of independent regions in different phases of their evolution from gas to stars. Our method measures the characteristic distance between these regions, which we denote as $\lambda_{\rm 21\mu m}$ and $\lambda_{\rm H\alpha}$ depending on which SFR tracer is being used. We find $\lambda_{\rm 21\mu m} = 90_{-17}^{+51} \: {\rm pc}$, showing a very good agreement with $\lambda_{\rm H\alpha} = 96_{-11}^{+13} \: {\rm pc}$. This $\lambda_{\rm 21\mu m}$ falls into the range of values found in five nearby galaxies using \textit{Spitzer} MIPS 24$\mu m$ as a SFR tracer \citep[70-200\,pc][]{kim22}, as well as that found in a larger sample of 54 galaxies using H$\alpha$ as a SFR tracer \citep[$100{-}400$\,pc;][]{chevance20, kim22}.

\section{Conclusion}
Using novel observations of NGC\,628 at a wavelength of 21\,$\mu$m from MIRI on JWST, together with CO from ALMA and narrow-band H$\alpha$ emission maps at matched resolution, we have characterized the evolutionary cycle of GMCs from their inert phase, to the onset of embedded massive star formation, the partially exposed star-forming phase, and finally to H~\textsc{ii} regions free of cold molecular gas. This is the first time that the start and the duration of the embedded phase of star formation can be characterized at a distance greater than 3.5\,Mpc, unlocking the necessary statistics and dynamic range for characterizing the environmental dependence of the physical processes driving the earliest phases of massive star formation.

We find that the time during which GMCs in NGC\,628 are truly free from massive star formation ($=t_{\rm CO}-t_{\rm fb, 21\,\mu m}$) is $18.8_{-3.6}^{+2.7}$\,Myr. The duration of the embedded phase of star formation ($t_{\rm fb, 21\mu m}$) is $5.1_{-1.4}^{+2.7}$\,Myr, representing $\sim$20\% of the cloud lifetime. The H$\alpha$ emission is heavily obscured during almost the entire first half of this phase, resulting in $t_{\rm obsc}$ of $2.3_{-1.4}^{+2.7}$\,Myr. Then, the star-forming region partially reveals itself from its natal GMC, causing the CO emission to be detected in association with 21\,$\mu$m and H$\alpha$ emission for a duration of $t_{\rm fb, H\alpha}=2.7_{-0.6}^{+0.5}$\,Myr. Finally, the molecular cloud is completely dispersed by stellar feedback, and only SFR tracers are detected for another $\sim$5\,Myr without associated CO emission.

\begin{figure}[t!]
\includegraphics[scale=0.47]{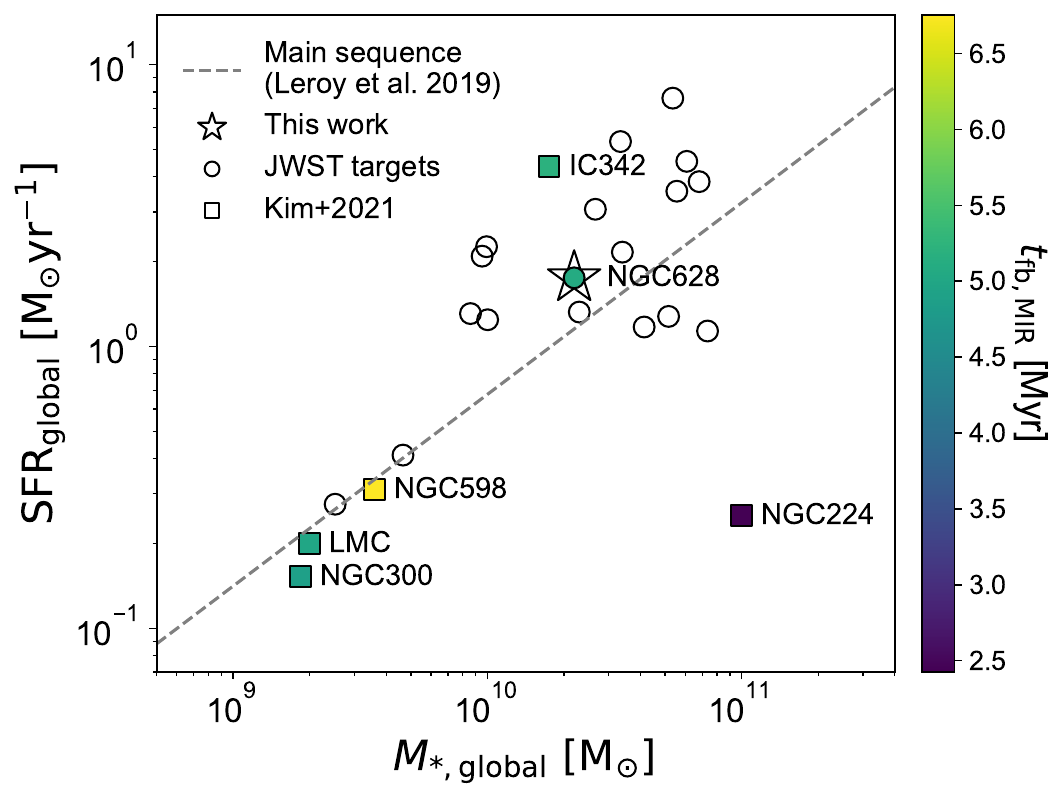}
\caption{Galaxy-wide SFR ($\rm SFR_{global}$) as a function of stellar mass ($M_{\rm *, global}$) for the PHANGS--JWST targets (circles), as well as nearby galaxies from \citet[squares;][]{kim21}. For NGC\,628 (this work) and nearby galaxies, the data points are colored by the duration of the embedded phase of star formation, derived using mid-infrared as a SFR tracer ($t_{\rm fb, MIR}$), which is JWST 21\,$\mu$m for NGC\, 0628 and \textit{Spitzer} 24\,$\mu$m for other galaxies. The dashed line shows the star-forming main-sequence of local galaxies \citep{leroy19}.} \label{fig:sample}
\end{figure}

In Figure~\ref{fig:sample}, the distribution of our PHANGS--JWST target galaxies as well as the five galaxies from \citet{kim22} are shown in the plane spanned by the galaxy stellar mass ($M_{\rm *, global}$) and the galaxy-wide SFR ($\rm SFR_{global}$). As a proof of concept, we have measured the timescales of the embedded and heavily obscured phases of star formation in one of the JWST target galaxies, NGC\,628. No trend is found between the duration of the embedded phase (here denoted by $t_{\rm fb, MIR}$ because the figure combines measurements from Spitzer at 24\,$\mu$m and from JWST at 21\,$\mu$m) and the galaxy properties (e.g.\ mass, SFR, offset from the main-sequence), but NGC\,628 represents an important extension of the parameter space shown here. Our results highlight the power of JWST by demonstrating that the quality of the data enables the embedded phase of star formation to be systematically characterized for a galaxy located at 9.8\,Mpc. With the arrival of JWST, the volume where such measurements can be done has increased by a factor of $>100$ (with $D<25$\,Mpc), compared to what was possible with \textit{Spitzer} (with $D<3.5$\,Mpc). Our measurements are in good agreement with those from our previous work in the small sample of five nearby galaxies (at $D<3.5$\,Mpc; \citealp{kim21}) for which such measurements were possible previously, and our results also achieve a comparable uncertainty of 30\%.

In the near future, a systematic determination of these timescales will become possible with the PHANGS--JWST survey, significantly increasing the total number of galaxies where this measurement can be performed to 24, where 19 of them come from PHANGS--JWST and have distances up to 20\,Mpc. This will for the first time cover a wide range of parameters (e.g.\ galaxy masses, morphological types, and ISM properties) across a statistically representative sample. Specifically, with the addition of the full PHANGS-JWST galaxy sample, the ranges of GMC properties where we can characterize this early phase of star formation become much wider. For example, the range of average internal pressure of GMCs in our previous galaxy sample \citep{kim21} was $10^4{-}10^5\,\rm K cm^{-3}$ and will be expanded up to $10^7\,\rm K cm^{-3}$. Similarly, the average molecular gas surface density was $10^{1}{-}10^{2}\,M_{\odot}\rm pc^{-2}$ and now can be probed up to  $10^{3}\,M_{\odot}\rm pc^{-2}$ (\citealp{rosolowsky21}; Hughes et al.\ in prep.).   This will allow us to characterize how the processes regulating the early stages of massive star formation depend on the galactic environment.

\section*{acknowledgments}
We thank an anonymous referee for helpful comments that improved the quality of the manuscript.
JK gratefully acknowledges funding from the Deutsche Forschungsgemeinschaft (DFG, German Research Foundation) through the DFG Sachbeihilfe (grant number KR4801/2-1).
MC gratefully acknowledges funding from the DFG through an Emmy Noether Research Group (grant number CH2137/1-1).
JMDK gratefully acknowledges funding from the DFG through an Emmy Noether Research Group (grant number KR4801/1-1), as well as from the European Research Council (ERC) under the European Union's Horizon 2020 research and innovation programme via the ERC Starting Grant MUSTANG (grant agreement number 714907). 
COOL Research DAO is a Decentralized Autonomous Organization supporting research in astrophysics aimed at uncovering our cosmic origins.
FB would like to acknowledge funding from the European Research Council (ERC) under the European Union’s Horizon 2020 research and innovation programme (grant agreement No.726384/Empire).
MB acknowledges support from FONDECYT regular grant 1211000 and by the ANID BASAL project FB210003.
E.C. acknowledges support from ANID Basal projects ACE210002 and FB210003.
OE, KK gratefully acknowledge funding from Deutsche Forschungsgemeinschaft (DFG, German Research Foundation) in the form of an Emmy Noether Research Group (grant number KR4598/2-1, PI Kreckel).
KG is supported by the Australian Research Council through the Discovery Early Career Researcher Award (DECRA) Fellowship DE220100766 funded by the Australian Government. KG is supported by the Australian Research Council Centre of Excellence for All Sky Astrophysics in 3 Dimensions (ASTRO~3D), through project number CE170100013. 
HH acknowledges the support of the Natural Sciences and Engineering Research Council of Canada (NSERC), funding reference number RGPIN-2022-03499.
AKL gratefully acknowledges support by grants 1653300 and 2205628 from the National Science Foundation, by award JWST-GO-02107.009-A, and by a Humboldt Research Award from the Alexander von Humboldt Foundation.
HAP acknowledges support by the National Science and Technology Council of Taiwan under grant 110-2112-M-032-020-MY3.
G.A.B. acknowledges the support from ANID Basal project FB210003.
JPe acknowledges support by the DAOISM grant ANR-21-CE31-0010 and by the Programme National ``Physique et Chimie du Milieu Interstellaire'' (PCMI) of CNRS/INSU with INC/INP, co-funded by CEA and CNES. 
MQ acknowledges support from the Spanish grant PID2019-106027GA-C44, funded by MCIN/AEI/10.13039/501100011033.
ES and TGW acknowledge funding from the European Research Council (ERC) under the European Union’s Horizon 2020 research and innovation programme (grant agreement No. 694343).
RJS acknowledges funding from an STFC ERF (grant ST/N00485X/1). 
EJW acknowledges the funding provided by the Deutsche Forschungsgemeinschaft (DFG, German Research Foundation) -- Project-ID 138713538 -- SFB 881 (``The Milky Way System'', subproject P1).

This work is based on observations made with the NASA/ESA/CSA James Webb Space Telescope. The data were obtained from the Mikulski Archive for Space Telescopes at the Space Telescope Science Institute, which is operated by the Association of Universities for Research in Astronomy, Inc., under NASA contract NAS 5-03127 for JWST. These observations are associated with program 2107.
This paper makes use of the following ALMA data: \linebreak
ADS/JAO.ALMA\#2012.1.00650.S\linebreak 
ALMA is a partnership of ESO (representing its member states), NSF (USA) and NINS (Japan), together with NRC (Canada), MOST and ASIAA (Taiwan), and KASI (Republic of Korea), in cooperation with the Republic of Chile. The Joint ALMA Observatory is operated by ESO, AUI/NRAO and NAOJ."
This paper includes data based on observations carried out at the MPG 2.2m telescope on La Silla, Chile.

%

\vspace{5mm}
\facilities{JWST (MIRI), ALMA, MPG--ESO:2.2m}


\software{astropy \citep{astropy}, numpy \citep{harris20}, scipy \citep{virtanen20}
          }



\appendix

\section{Accuracy of our measurements\label{app:acc}}

\citet[][sect. 4.4]{kruijssen18} have outlined a set of criteria that our measurements have to satisfy in order to be considered reliable with an accuracy of 30\% or better. Here, we verify that these requirements are fulfilled, demonstrating that the constrained $t_{\rm 21 \mu m}$, $t_{\rm fb, 21 \mu m}$, and $\lambda_{\rm 21 \mu m}$ are accurate. We refer to our previous papers \citep{chevance20, chevance22, kim22} for a validation of our measurements using H$\alpha$ as a SFR tracer. 

\begin{enumerate}
\item The emitting timescale of molecular gas and SFR tracer should not differ by one order of magnitude. This is satisfied by $|\log_{10}(t_{\rm 21\mu m}/t_{\rm CO})| = 0.35$.
\item Individual regions within a galaxy should be sufficiently resolved and this is ensured by $\lambda_{\rm 21 \mu m}/l_{\rm ap, min}=1.9$.
\item We confirm that the number of identified peaks in each CO and 21\,$\mu$m map is more than 35 peaks and is $\sim400$ on average.
\item The CO-to-21\,$\mu$m flux ratio measured locally when focusing on CO (resp.~21\,$\mu$m) peaks should not fall below (resp.~above) the galactic average. This is visibly satisfied in Figure~\ref{fig:result} and confirms that any diffuse, large-scale emission has been appropriately filtered in both maps.
\item In order to ensure that the identified peaks represent a temporal manifestation of regions undergoing independent evolution from gas to stars, the galaxy-wide SFR during the last GMC cycle ($\tau=t_{\rm CO}+t_{\rm 21 \mu m}-t_{\rm fb, \rm 21 \mu m}=27.5$\,Myr) should not vary by more than 0.2\,dex, when averaged over a bin size of $t_{\rm CO}$ or $t_{\rm 21 \mu m}$. This is confirmed by the star formation history derived using MUSE data and spectral fitting by I.~Pessa et al.\ (in preparation), where the SFR is found to not vary significantly during the most recent $\sim$30\,Myr, when time averaged by $t_{\rm 21\mu m}\approx 10$\,Myr. 
\item Individual regions should be observable in both molecular gas and SFR tracer at some point in their evolution. This implies that the CO and 21\,$\mu$m maps should be sensitive to similar regions. In order to confirm this, we first compute the minimum mass of the young stellar population that is expected to form within the observed clouds, by multiplying the 5$\sigma$ sensitivity of the CO map ($5\times10^{4}\,M_{\odot}$ by the integrated star formation efficiency ($5.5_{-2.3}^{+4.0}$\%), measured for clouds in NGC\,628 \citep{kim22}. Then, this value is compared to the mass of a hypothetical young stellar population that emits photons at the 5$\sigma$ sensitivity of the 21\,$\mu$m map on the scale of star-forming regions ($\lambda_{21\,\mu m}$). We use the \textsc{Starburst99} model \citep{leitherer99} to estimate H$\alpha$ luminosity, which is converted to 21\,$\mu$m using the conversion factor from \citet{LEROY1_PHANGSJWST} to estimate the mass, assuming instanteneous star formation, 5\,Myr ago. As a result, we find a reasonable agreement of the expected minimum mass of the stellar population between that obtained from CO map ($\sim 2500\,M_{\odot}$) and that from 21\,$\mu$m map ($\sim 1500\,M_{\odot}$). However, we note that the resolution and sensitivity of the ALMA map may quickly become the limiting factor to these measurements using JWST observations.
\begin{figure}
\includegraphics[scale=0.7]{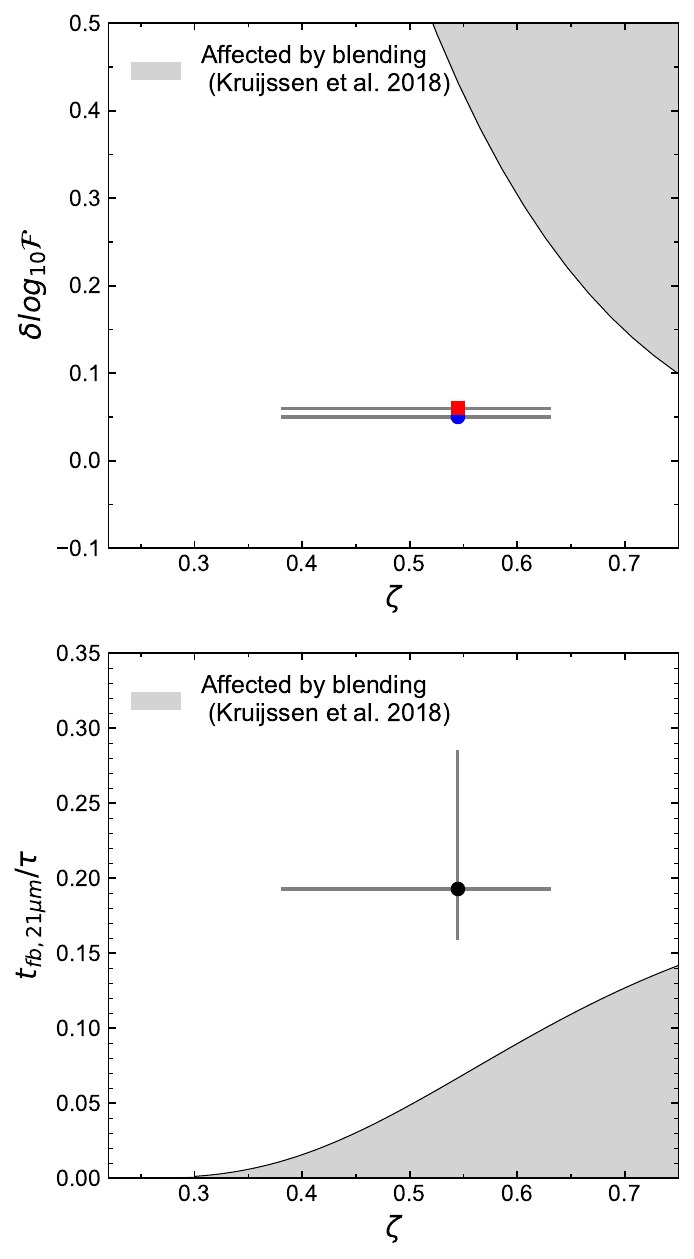}
\caption{The top panel shows the flux contrast ($\rm{\delta}log_{10}\mathcal{F}$) used to identify peaks on CO (blue) 21\,$\mu$m (red) map as a function of the average filling factor $\zeta$. The bottom panel shows the ratio of the feedback timescale ($t_{\rm fb, 21\mu m}$) and the total duration of the GMC lifecycle ($\tau$) as a function of $\zeta$. In both panels, the shaded region shows the parameter space where crowding of sources can lead to an overestimation of the feedback timescale. Our data points are well outside of this region, confirming that we sufficiently resolve star-forming regions and our measurement of $t_{\rm fb, 21\,\mu m}$ is reliable.   } \label{fig:blending}
\end{figure}

\item When peaks are crowded and potentially overlapping with each other, the flux contrast used for peak identification ($\rm{\delta}log_{10}\mathcal{F}$) should be small enough to pick out adjacent peaks, and avoid overestimating the feedback timescale. \citet{kruijssen18} have prescribed an upper limit of this value as a function of the average filling factor of gas and SFR tracer peaks ($\zeta$). This $\zeta$ is defined as $2r/\lambda$, where $r$ is the mean radius of the peaks of a given tracer. The total $\zeta$ is obtained by averaging the filling factor for the gas and SFR tracer peaks, weighted by their associated emission timescales. In Figure~\ref{fig:blending}, we show that our selection of $\rm{\delta}log_{10}\mathcal{F}$ for both CO and 21\,$\mu$m is below the upper limit determined by \citet{kruijssen18}.
\item  Even when the previous condition is met, peaks can be overlapping with neighbouring peaks due to high filling factors, and this can falsely cause a longer feedback timescale to be measured. In this case, the measured feedback timescale would only be an upper limit. In Figure~\ref{fig:blending}, we compare our measurements of $t_{\rm fb, 21\mu m}/\tau$ and $\zeta$ to the analytic prescription by \citet{kruijssen18}, in which the shaded area represents the parameter space where crowding of peaks are affecting our measurements of the feedback timescale. Our measurements are well outside of this shaded region, indicating that peaks are sufficiently resolved.
\item We confirm that the conditions $t_{\rm fb} > 0.05\,\tau$ and $t_{\rm fb} < 0.95\, \tau$ are satisfied by $t_{\rm fb} \approx 0.2 \,\tau$, as shown in the lower panel of Figure~\ref{fig:blending}.
\item With a similar reasoning as for condition 5, we do not expect the galaxy-wide SFR to vary more than 0.2\,dex during the last course of $\tau$, when time-averaged over the feedback timescale.
\item After masking the crowded galactic center, we confirm that visual inspection does not reveal regions with abundant blending. 

\end{enumerate}

\section{Data used in Figure~\lowercase{\ref{fig:result}}\label{app:data}}

The left panel of Figure~\ref{fig:result} shows the measured deviation of CO-to-21\,$\mu$m flux ratio relative to the galactic average, as a function of the size of apertures centered on CO and 21\,$\mu$m emission peaks. The measured flux ratios increasingly diverge from the galactic average value towards smaller scales, illustrating that molecular gas and young stellar regions are spatially decorrelated. Table~\ref{tab:app_result} lists the measured flux ratios as a function of the aperture size used to make the left panel of Figure~\ref{fig:result}.

\begin{table*}[!b]
\begin{center}
\caption{Data used in Figure~\ref{fig:result}. Relative changes of the CO-to-21\,$\mu$m flux ratio compared to the galactic average as a function of the size of apertures focused on CO and 21\,$\mu$m emission peaks. The downward and upward 1$\sigma$ uncertainties of each measurement ($\sigma_{\rm min}$ and $\sigma_{\rm max}$), as well as those accounting for the covariance between data points are listed ($\sigma_{\rm min}^{\rm shade}$ and $\sigma_{\rm max}^{\rm shade}$). \label{tab:app_result}}
\setlength\tabcolsep{8pt}
\begin{tabular}{ccccccccccc}
\hline
Aperture &  Centered on&  $\sigma_{\rm min}$& $\sigma_{\rm max}$ &  $\sigma_{\rm min}^{\rm shade}$& $\sigma_{\rm max}^{\rm shade}$  &  Centered on&  $\sigma_{\rm min}$& $\sigma_{\rm max}$ &  $\sigma_{\rm min}^{\rm shade}$& $\sigma_{\rm max}^{\rm shade}$ \\
size [pc] &  CO peaks &  &&&&  21\,$\mu$m peaks &&&&\\ 
\hline
50&1.37&0.16&0.18&0.03&0.03&0.41&0.03&0.04&0.01&0.01\\
63&1.33&0.15&0.17&0.03&0.03&0.43&0.04&0.04&0.01&0.01\\
81&1.35&0.15&0.17&0.04&0.04&0.47&0.04&0.04&0.01&0.01\\
103&1.29&0.14&0.16&0.04&0.04&0.51&0.04&0.05&0.01&0.01\\
131&1.22&0.13&0.14&0.03&0.03&0.60&0.06&0.06&0.01&0.01\\
166&1.22&0.13&0.15&0.03&0.04&0.67&0.06&0.07&0.02&0.02\\
211&1.20&0.13&0.15&0.04&0.04&0.77&0.08&0.09&0.02&0.02\\
268&1.18&0.13&0.15&0.03&0.04&0.82&0.08&0.09&0.02&0.02\\
340&1.13&0.12&0.14&0.03&0.03&0.85&0.09&0.10&0.02&0.02\\
431&1.07&0.11&0.12&0.03&0.03&0.86&0.09&0.10&0.02&0.02\\
544&1.05&0.11&0.12&0.03&0.03&0.90&0.09&0.11&0.02&0.02\\
690&1.03&0.10&0.11&0.02&0.03&0.96&0.09&0.11&0.02&0.02\\
866&1.04&0.10&0.11&0.02&0.02&0.98&0.10&0.11&0.02&0.02\\
1088&1.04&0.10&0.11&0.02&0.02&1.01&0.10&0.11&0.02&0.02\\
1360&1.02&0.10&0.11&0.02&0.02&1.00&0.10&0.11&0.02&0.02\\
\hline
\end{tabular}
\end{center}
\end{table*}

\bibliographystyle{aasjournal}
\bibliography{mybib, phangsjwst}{} 

\begin{thebibliography}{}
\expandafter\ifx\csname natexlab\endcsname\relax\def\natexlab#1{#1}\fi
\providecommand{\url}[1]{\href{#1}{#1}}
\providecommand{\dodoi}[1]{doi:~\href{http://doi.org/#1}{\nolinkurl{#1}}}
\providecommand{\doeprint}[1]{\href{http://ascl.net/#1}{\nolinkurl{http://ascl.net/#1}}}
\providecommand{\doarXiv}[1]{\href{https://arxiv.org/abs/#1}{\nolinkurl{https://arxiv.org/abs/#1}}}

\bibitem[{{Anand} {et~al.}(2021{\natexlab{a}}){Anand}, {Lee}, {Van Dyk},
  {Leroy}, {Rosolowsky}, {Schinnerer}, {Larson}, {Kourkchi}, {Kreckel},
  {Scheuermann}, {Rizzi}, {Thilker}, {Tully}, {Bigiel}, {Blanc}, {Boquien},
  {Chandar}, {Dale}, {Emsellem}, {Deger}, {Glover}, {Grasha}, {Groves}, {S.
  Klessen}, {Kruijssen}, {Querejeta}, {S{\'a}nchez-Bl{\'a}zquez}, {Schruba},
  {Turner}, {Ubeda}, {Williams}, \& {Whitmore}}]{anand21}
{Anand}, G.~S., {Lee}, J.~C., {Van Dyk}, S.~D., {et~al.} 2021{\natexlab{a}},
  \mnras, 501, 3621, \dodoi{10.1093/mnras/staa3668}

\bibitem[{{Anand} {et~al.}(2021{\natexlab{b}}){Anand}, {Rizzi}, {Tully},
  {Shaya}, {Karachentsev}, {Makarov}, {Makarova}, {Wu}, {Dolphin}, \&
  {Kourkchi}}]{anand21_2}
{Anand}, G.~S., {Rizzi}, L., {Tully}, R.~B., {et~al.} 2021{\natexlab{b}}, \aj,
  162, 80, \dodoi{10.3847/1538-3881/ac0440}

\bibitem[{{Aniano} {et~al.}(2011){Aniano}, {Draine}, {Gordon}, \&
  {Sandstrom}}]{aniano11}
{Aniano}, G., {Draine}, B.~T., {Gordon}, K.~D., \& {Sandstrom}, K. 2011, \pasp,
  123, 1218, \dodoi{10.1086/662219}

\bibitem[{{Astropy Collaboration} {et~al.}(2013){Astropy Collaboration},
  {Robitaille}, {Tollerud}, {Greenfield}, {Droettboom}, {Bray}, {Aldcroft},
  {Davis}, {Ginsburg}, {Price-Whelan}, {Kerzendorf}, {Conley}, {Crighton},
  {Barbary}, {Muna}, {Ferguson}, {Grollier}, {Parikh}, {Nair}, {Unther},
  {Deil}, {Woillez}, {Conseil}, {Kramer}, {Turner}, {Singer}, {Fox}, {Weaver},
  {Zabalza}, {Edwards}, {Azalee Bostroem}, {Burke}, {Casey}, {Crawford},
  {Dencheva}, {Ely}, {Jenness}, {Labrie}, {Lim}, {Pierfederici}, {Pontzen},
  {Ptak}, {Refsdal}, {Servillat}, \& {Streicher}}]{astropy}
{Astropy Collaboration}, {Robitaille}, T.~P., {Tollerud}, E.~J., {et~al.} 2013,
  \aap, 558, A33, \dodoi{10.1051/0004-6361/201322068}

\bibitem[{{Backus} {et~al.}(2005){Backus}, {Velusamy}, {Thompson}, \&
  {Arballo}}]{backus05}
{Backus}, C., {Velusamy}, T., {Thompson}, T., \& {Arballo}, J. 2005,
  Astronomical Society of the Pacific Conference Series, Vol. 347, {Hires:
  Super-resolution for the Spitzer Space Telescope}, ed. P.~{Shopbell},
  M.~{Britton}, \& R.~{Ebert}, 61

\bibitem[{{Barnes} {et~al.}(2020){Barnes}, {Longmore}, {Dale}, {Krumholz},
  {Kruijssen}, \& {Bigiel}}]{barnes20}
{Barnes}, A.~T., {Longmore}, S.~N., {Dale}, J.~E., {et~al.} 2020, \mnras, 498,
  4906, \dodoi{10.1093/mnras/staa2719}

\bibitem[{{Barnes} {et~al.}(2022){Barnes}, {Chandar}, {Kreckel}, {Glover},
  {Scheuermann}, {Belfiore}, {Bigiel}, {Blanc}, {Boquien}, {den Brok},
  {Congiu}, {Chevance}, {Dale}, {Deger}, {Kruijssen}, {Egorov}, {Eibensteiner},
  {Emsellem}, {Grasha}, {Groves}, {Klessen}, {Hannon}, {Hassani}, {Lee},
  {Leroy}, {Lopez}, {McLeod}, {Pan}, {S{\'a}nchez-Bl{\'a}zquez}, {Schinnerer},
  {Sormani}, {Thilker}, {Ubeda}, {Watkins}, \& {Williams}}]{barnes22}
{Barnes}, A.~T., {Chandar}, R., {Kreckel}, K., {et~al.} 2022, \aap, 662, L6,
  \dodoi{10.1051/0004-6361/202243766}

\bibitem[{{Battersby} {et~al.}(2017){Battersby}, {Bally}, \&
  {Svoboda}}]{battersby17}
{Battersby}, C., {Bally}, J., \& {Svoboda}, B. 2017, \apj, 835, 263,
  \dodoi{10.3847/1538-4357/835/2/263}

\bibitem[{{Bigiel} {et~al.}(2008){Bigiel}, {Leroy}, {Walter}, {Brinks}, {de
  Blok}, {Madore}, \& {Thornley}}]{bigiel08}
{Bigiel}, F., {Leroy}, A., {Walter}, F., {et~al.} 2008, \aj, 136, 2846,
  \dodoi{10.1088/0004-6256/136/6/2846}

\bibitem[{{Blitz} {et~al.}(2007){Blitz}, {Fukui}, {Kawamura}, {Leroy},
  {Mizuno}, \& {Rosolowsky}}]{blitz07}
{Blitz}, L., {Fukui}, Y., {Kawamura}, A., {et~al.} 2007, in Protostars and
  Planets V, ed. B.~{Reipurth}, D.~{Jewitt}, \& K.~{Keil}, 81.
\newblock \doarXiv{astro-ph/0602600}

\bibitem[{{Calzetti} {et~al.}(2015){Calzetti}, {Johnson}, {Adamo}, {Gallagher},
  {Andrews}, {Smith}, {Clayton}, {Lee}, {Sabbi}, {Ubeda}, {Kim}, {Ryon},
  {Thilker}, {Bright}, {Zackrisson}, {Kennicutt}, {de Mink}, {Whitmore},
  {Aloisi}, {Chandar}, {Cignoni}, {Cook}, {Dale}, {Elmegreen}, {Elmegreen},
  {Evans}, {Fumagalli}, {Gouliermis}, {Grasha}, {Grebel}, {Krumholz},
  {Walterbos}, {Wofford}, {Brown}, {Christian}, {Dobbs}, {Herrero}, {Kahre},
  {Messa}, {Nair}, {Nota}, {{\"O}stlin}, {Pellerin}, {Sacchi}, {Schaerer}, \&
  {Tosi}}]{calzetti15}
{Calzetti}, D., {Johnson}, K.~E., {Adamo}, A., {et~al.} 2015, \apj, 811, 75,
  \dodoi{10.1088/0004-637X/811/2/75}

\bibitem[{{Chevance} {et~al.}(2022{\natexlab{a}}){Chevance}, {Krumholz},
  {McLeod}, {Ostriker}, {Rosolowsky}, \& {Sternberg}}]{chevance22_rev}
{Chevance}, M., {Krumholz}, M.~R., {McLeod}, A.~F., {et~al.}
  2022{\natexlab{a}}, arXiv e-prints, arXiv:2203.09570.
\newblock \doarXiv{2203.09570}

\bibitem[{{Chevance} {et~al.}(2020{\natexlab{a}}){Chevance}, {Kruijssen},
  {Vazquez-Semadeni}, {Nakamura}, {Klessen}, {Ballesteros-Paredes}, {Inutsuka},
  {Adamo}, \& {Hennebelle}}]{chevance20_rev}
{Chevance}, M., {Kruijssen}, J.~M.~D., {Vazquez-Semadeni}, E., {et~al.}
  2020{\natexlab{a}}, \ssr, 216, 50, \dodoi{10.1007/s11214-020-00674-x}

\bibitem[{{Chevance} {et~al.}(2020{\natexlab{b}}){Chevance}, {Kruijssen},
  {Hygate}, {Schruba}, {Longmore}, {Groves}, {Henshaw}, {Herrera}, {Hughes},
  {Jeffreson}, {Lang}, {Leroy}, {Meidt}, {Pety}, {Razza}, {Rosolowsky},
  {Schinnerer}, {Bigiel}, {Blanc}, {Emsellem}, {Faesi}, {Glover}, {Haydon},
  {Ho}, {Kreckel}, {Lee}, {Liu}, {Querejeta}, {Saito}, {Sun}, {Usero}, \&
  {Utomo}}]{chevance20}
{Chevance}, M., {Kruijssen}, J.~M.~D., {Hygate}, A. P.~S., {et~al.}
  2020{\natexlab{b}}, \mnras, 493, 2872, \dodoi{10.1093/mnras/stz3525}

\bibitem[{{Chevance} {et~al.}(2022{\natexlab{b}}){Chevance}, {Kruijssen},
  {Krumholz}, {Groves}, {Keller}, {Hughes}, {Glover}, {Henshaw}, {Herrera},
  {Kim}, {Leroy}, {Pety}, {Razza}, {Rosolowsky}, {Schinnerer}, {Schruba},
  {Barnes}, {Bigiel}, {Blanc}, {Dale}, {Emsellem}, {Faesi}, {Grasha},
  {Klessen}, {Kreckel}, {Liu}, {Longmore}, {Meidt}, {Querejeta}, {Saito},
  {Sun}, \& {Usero}}]{chevance22}
{Chevance}, M., {Kruijssen}, J.~M.~D., {Krumholz}, M.~R., {et~al.}
  2022{\natexlab{b}}, \mnras, 509, 272, \dodoi{10.1093/mnras/stab2938}

\bibitem[{{Corbelli} {et~al.}(2017){Corbelli}, {Braine}, {Bandiera},
  {Brouillet}, {Combes}, {Druard}, {Gratier}, {Mata}, {Schuster}, {Xilouris},
  \& {Palla}}]{corbelli17}
{Corbelli}, E., {Braine}, J., {Bandiera}, R., {et~al.} 2017, \aap, 601, A146,
  \dodoi{10.1051/0004-6361/201630034}

\bibitem[{{Della Bruna} {et~al.}(2022){Della Bruna}, {Adamo}, {McLeod},
  {Smith}, {Savard}, {Robert}, {Sun}, {Amram}, {Bik}, {Blair}, {Long},
  {Renaud}, {Walterbos}, \& {Usher}}]{dellabruna22}
{Della Bruna}, L., {Adamo}, A., {McLeod}, A.~F., {et~al.} 2022, \aap, 666, A29,
  \dodoi{10.1051/0004-6361/202243395}

\bibitem[{{Draine} \& {Li}(2007)}]{draine07}
{Draine}, B.~T., \& {Li}, A. 2007, \apj, 657, 810, \dodoi{10.1086/511055}

\bibitem[{{Dumas} {et~al.}(2011){Dumas}, {Schinnerer}, {Tabatabaei}, {Beck},
  {Velusamy}, \& {Murphy}}]{dumas11}
{Dumas}, G., {Schinnerer}, E., {Tabatabaei}, F.~S., {et~al.} 2011, \aj, 141,
  41, \dodoi{10.1088/0004-6256/141/2/41}

\bibitem[{{Engargiola} {et~al.}(2003){Engargiola}, {Plambeck}, {Rosolowsky}, \&
  {Blitz}}]{engargiola03}
{Engargiola}, G., {Plambeck}, R.~L., {Rosolowsky}, E., \& {Blitz}, L. 2003,
  \apjs, 149, 343, \dodoi{10.1086/379165}

\bibitem[{{Gaia Collaboration} {et~al.}(2018){Gaia Collaboration}, {Brown},
  {Vallenari}, {Prusti}, {de Bruijne}, {Babusiaux}, {Bailer-Jones}, {Biermann},
  {Evans}, {Eyer}, {Jansen}, {Jordi}, {Klioner}, {Lammers}, {Lindegren},
  {Luri}, {Mignard}, {Panem}, {Pourbaix}, {Randich}, {Sartoretti}, {Siddiqui},
  {Soubiran}, {van Leeuwen}, {Walton}, {Arenou}, {Bastian}, {Cropper},
  {Drimmel}, {Katz}, {Lattanzi}, {Bakker}, {Cacciari}, {Casta{\~n}eda},
  {Chaoul}, {Cheek}, {De Angeli}, {Fabricius}, {Guerra}, {Holl}, {Masana},
  {Messineo}, {Mowlavi}, {Nienartowicz}, {Panuzzo}, {Portell}, {Riello},
  {Seabroke}, {Tanga}, {Th{\'e}venin}, {Gracia-Abril}, {Comoretto},
  {Garcia-Reinaldos}, {Teyssier}, {Altmann}, {Andrae}, {Audard},
  {Bellas-Velidis}, {Benson}, {Berthier}, {Blomme}, {Burgess}, {Busso},
  {Carry}, {Cellino}, {Clementini}, {Clotet}, {Creevey}, {Davidson}, {De
  Ridder}, {Delchambre}, {Dell'Oro}, {Ducourant},
  {Fern{\'a}ndez-Hern{\'a}ndez}, {Fouesneau}, {Fr{\'e}mat}, {Galluccio},
  {Garc{\'\i}a-Torres}, {Gonz{\'a}lez-N{\'u}{\~n}ez}, {Gonz{\'a}lez-Vidal},
  {Gosset}, {Guy}, {Halbwachs}, {Hambly}, {Harrison}, {Hern{\'a}ndez},
  {Hestroffer}, {Hodgkin}, {Hutton}, {Jasniewicz}, {Jean-Antoine-Piccolo},
  {Jordan}, {Korn}, {Krone-Martins}, {Lanzafame}, {Lebzelter}, {L{\"o}ffler},
  {Manteiga}, {Marrese}, {Mart{\'\i}n-Fleitas}, {Moitinho}, {Mora}, {Muinonen},
  {Osinde}, {Pancino}, {Pauwels}, {Petit}, {Recio-Blanco}, {Richards},
  {Rimoldini}, {Robin}, {Sarro}, {Siopis}, {Smith}, {Sozzetti}, {S{\"u}veges},
  {Torra}, {van Reeven}, {Abbas}, {Abreu Aramburu}, {Accart}, {Aerts},
  {Altavilla}, {{\'A}lvarez}, {Alvarez}, {Alves}, {Anderson}, {Andrei},
  {Anglada Varela}, {Antiche}, {Antoja}, {Arcay}, {Astraatmadja}, {Bach},
  {Baker}, {Balaguer-N{\'u}{\~n}ez}, {Balm}, {Barache}, {Barata}, {Barbato},
  {Barblan}, {Barklem}, {Barrado}, {Barros}, {Barstow}, {Bartholom{\'e}
  Mu{\~n}oz}, {Bassilana}, {Becciani}, {Bellazzini}, {Berihuete}, {Bertone},
  {Bianchi}, {Bienaym{\'e}}, {Blanco-Cuaresma}, {Boch}, {Boeche}, {Bombrun},
  {Borrachero}, {Bossini}, {Bouquillon}, {Bourda}, {Bragaglia}, {Bramante},
  {Breddels}, {Bressan}, {Brouillet}, {Br{\"u}semeister}, {Brugaletta},
  {Bucciarelli}, {Burlacu}, {Busonero}, {Butkevich}, {Buzzi}, {Caffau},
  {Cancelliere}, {Cannizzaro}, {Cantat-Gaudin}, {Carballo}, {Carlucci},
  {Carrasco}, {Casamiquela}, {Castellani}, {Castro-Ginard}, {Charlot},
  {Chemin}, {Chiavassa}, {Cocozza}, {Costigan}, {Cowell}, {Crifo}, {Crosta},
  {Crowley}, {Cuypers}, {Dafonte}, {Damerdji}, {Dapergolas}, {David}, {David},
  {de Laverny}, {De Luise}, {De March}, {de Martino}, {de Souza}, {de Torres},
  {Debosscher}, {del Pozo}, {Delbo}, {Delgado}, {Delgado}, {Di Matteo},
  {Diakite}, {Diener}, {Distefano}, {Dolding}, {Drazinos}, {Dur{\'a}n},
  {Edvardsson}, {Enke}, {Eriksson}, {Esquej}, {Eynard Bontemps}, {Fabre},
  {Fabrizio}, {Faigler}, {Falc{\~a}o}, {Farr{\`a}s Casas}, {Federici},
  {Fedorets}, {Fernique}, {Figueras}, {Filippi}, {Findeisen}, {Fonti},
  {Fraile}, {Fraser}, {Fr{\'e}zouls}, {Gai}, {Galleti}, {Garabato},
  {Garc{\'\i}a-Sedano}, {Garofalo}, {Garralda}, {Gavel}, {Gavras}, {Gerssen},
  {Geyer}, {Giacobbe}, {Gilmore}, {Girona}, {Giuffrida}, {Glass}, {Gomes},
  {Granvik}, {Gueguen}, {Guerrier}, {Guiraud}, {Guti{\'e}rrez-S{\'a}nchez},
  {Haigron}, {Hatzidimitriou}, {Hauser}, {Haywood}, {Heiter}, {Helmi}, {Heu},
  {Hilger}, {Hobbs}, {Hofmann}, {Holland}, {Huckle}, {Hypki}, {Icardi},
  {Jan{\ss}en}, {Jevardat de Fombelle}, {Jonker}, {Juh{\'a}sz}, {Julbe},
  {Karampelas}, {Kewley}, {Klar}, {Kochoska}, {Kohley}, {Kolenberg},
  {Kontizas}, {Kontizas}, {Koposov}, {Kordopatis}, {Kostrzewa-Rutkowska},
  {Koubsky}, {Lambert}, {Lanza}, {Lasne}, {Lavigne}, {Le Fustec}, {Le
  Poncin-Lafitte}, {Lebreton}, {Leccia}, {Leclerc}, {Lecoeur-Taibi},
  {Lenhardt}, {Leroux}, {Liao}, {Licata}, {Lindstr{\o}m}, {Lister}, {Livanou},
  {Lobel}, {L{\'o}pez}, {Managau}, {Mann}, {Mantelet}, {Marchal}, {Marchant},
  {Marconi}, {Marinoni}, {Marschalk{\'o}}, {Marshall}, {Martino}, {Marton},
  {Mary}, {Massari}, {Matijevi{\v{c}}}, {Mazeh}, {McMillan}, {Messina},
  {Michalik}, {Millar}, {Molina}, {Molinaro}, {Moln{\'a}r}, {Montegriffo},
  {Mor}, {Morbidelli}, {Morel}, {Morris}, {Mulone}, {Muraveva}, {Musella},
  {Nelemans}, {Nicastro}, {Noval}, {O'Mullane}, {Ord{\'e}novic},
  {Ord{\'o}{\~n}ez-Blanco}, {Osborne}, {Pagani}, {Pagano}, {Pailler},
  {Palacin}, {Palaversa}, {Panahi}, {Pawlak}, {Piersimoni}, {Pineau}, {Plachy},
  {Plum}, {Poggio}, {Poujoulet}, {Pr{\v{s}}a}, {Pulone}, {Racero}, {Ragaini},
  {Rambaux}, {Ramos-Lerate}, {Regibo}, {Reyl{\'e}}, {Riclet}, {Ripepi}, {Riva},
  {Rivard}, {Rixon}, {Roegiers}, {Roelens}, {Romero-G{\'o}mez}, {Rowell},
  {Royer}, {Ruiz-Dern}, {Sadowski}, {Sagrist{\`a} Sell{\'e}s}, {Sahlmann},
  {Salgado}, {Salguero}, {Sanna}, {Santana-Ros}, {Sarasso}, {Savietto},
  {Schultheis}, {Sciacca}, {Segol}, {Segovia}, {S{\'e}gransan}, {Shih},
  {Siltala}, {Silva}, {Smart}, {Smith}, {Solano}, {Solitro}, {Sordo}, {Soria
  Nieto}, {Souchay}, {Spagna}, {Spoto}, {Stampa}, {Steele},
  {Steidelm{\"u}ller}, {Stephenson}, {Stoev}, {Suess}, {Surdej}, {Szabados},
  {Szegedi-Elek}, {Tapiador}, {Taris}, {Tauran}, {Taylor}, {Teixeira},
  {Terrett}, {Teyssand ier}, {Thuillot}, {Titarenko}, {Torra Clotet}, {Turon},
  {Ulla}, {Utrilla}, {Uzzi}, {Vaillant}, {Valentini}, {Valette}, {van Elteren},
  {Van Hemelryck}, {van Leeuwen}, {Vaschetto}, {Vecchiato}, {Veljanoski},
  {Viala}, {Vicente}, {Vogt}, {von Essen}, {Voss}, {Votruba}, {Voutsinas},
  {Walmsley}, {Weiler}, {Wertz}, {Wevers}, {Wyrzykowski}, {Yoldas},
  {{\v{Z}}erjal}, {Ziaeepour}, {Zorec}, {Zschocke}, {Zucker}, {Zurbach}, \&
  {Zwitter}}]{gaiadr2}
{Gaia Collaboration}, {Brown}, A.~G.~A., {Vallenari}, A., {et~al.} 2018, \aap,
  616, A1, \dodoi{10.1051/0004-6361/201833051}

\bibitem[{{Galliano} {et~al.}(2018){Galliano}, {Galametz}, \&
  {Jones}}]{galliano18}
{Galliano}, F., {Galametz}, M., \& {Jones}, A.~P. 2018, \araa, 56, 673,
  \dodoi{10.1146/annurev-astro-081817-051900}

\bibitem[{{Ginsburg} {et~al.}(2012){Ginsburg}, {Bressert}, {Bally}, \&
  {Battersby}}]{ginsburg12}
{Ginsburg}, A., {Bressert}, E., {Bally}, J., \& {Battersby}, C. 2012, \apjl,
  758, L29, \dodoi{10.1088/2041-8205/758/2/L29}

\bibitem[{{Grasha} {et~al.}(2018){Grasha}, {Calzetti}, {Bittle}, {Johnson},
  {Donovan Meyer}, {Kennicutt}, {Elmegreen}, {Adamo}, {Krumholz}, {Fumagalli},
  {Grebel}, {Gouliermis}, {Cook}, {Gallagher}, {Aloisi}, {Dale}, {Linden},
  {Sacchi}, {Thilker}, {Walterbos}, {Messa}, {Wofford}, \& {Smith}}]{grasha18}
{Grasha}, K., {Calzetti}, D., {Bittle}, L., {et~al.} 2018, \mnras, 481, 1016,
  \dodoi{10.1093/mnras/sty2154}

\bibitem[{{Grasha} {et~al.}(2019){Grasha}, {Calzetti}, {Adamo}, {Kennicutt},
  {Elmegreen}, {Messa}, {Dale}, {Fedorenko}, {Mahadevan}, {Grebel},
  {Fumagalli}, {Kim}, {Dobbs}, {Gouliermis}, {Ashworth}, {Gallagher}, {Smith},
  {Tosi}, {Whitmore}, {Schinnerer}, {Colombo}, {Hughes}, {Leroy}, \&
  {Meidt}}]{grasha19}
{Grasha}, K., {Calzetti}, D., {Adamo}, A., {et~al.} 2019, \mnras, 483, 4707,
  \dodoi{10.1093/mnras/sty3424}

\bibitem[{{Hannon} {et~al.}(2019){Hannon}, {Lee}, {Whitmore}, {Chandar},
  {Adamo}, {Mobasher}, {Aloisi}, {Calzetti}, {Cignoni}, {Cook}, {Dale},
  {Deger}, {Della Bruna}, {Elmegreen}, {Gouliermis}, {Grasha}, {Grebel},
  {Herrero}, {Hunter}, {Johnson}, {Kennicutt}, {Kim}, {Sacchi}, {Smith},
  {Thilker}, {Turner}, {Walterbos}, \& {Wofford}}]{hannon19}
{Hannon}, S., {Lee}, J.~C., {Whitmore}, B.~C., {et~al.} 2019, \mnras, 490,
  4648, \dodoi{10.1093/mnras/stz2820}

\bibitem[{{Hannon} {et~al.}(2022){Hannon}, {Lee}, {Whitmore}, {Mobasher},
  {Thilker}, {Chandar}, {Adamo}, {Wofford}, {Orozco-Duarte}, {Calzetti}, {Della
  Bruna}, {Kreckel}, {Groves}, {Barnes}, {Boquien}, {Belfiore}, \&
  {Linden}}]{hannon22}
---. 2022, \mnras, 512, 1294, \dodoi{10.1093/mnras/stac550}

\bibitem[{Harris {et~al.}(2020)Harris, Millman, van~der Walt, Gommers,
  Virtanen, Cournapeau, Wieser, Taylor, Berg, Smith, Kern, Picus, Hoyer, van
  Kerkwijk, Brett, Haldane, del R{\'{i}}o, Wiebe, Peterson,
  G{\'{e}}rard-Marchant, Sheppard, Reddy, Weckesser, Abbasi, Gohlke, \&
  Oliphant}]{harris20}
Harris, C.~R., Millman, K.~J., van~der Walt, S.~J., {et~al.} 2020, Nature, 585,
  357, \dodoi{10.1038/s41586-020-2649-2}

\bibitem[{{Hassani} {et~al.}(subm.)}]{HASSANI_PHANGSJWST}
{Hassani}, H., {et~al.} subm., \apjl

\bibitem[{{Haydon} {et~al.}(2020){Haydon}, {Kruijssen}, {Chevance}, {Hygate},
  {Krumholz}, {Schruba}, \& {Longmore}}]{haydon20}
{Haydon}, D.~T., {Kruijssen}, J.~M.~D., {Chevance}, M., {et~al.} 2020, \mnras,
  498, 235, \dodoi{10.1093/mnras/staa2430}

\bibitem[{{Hygate} {et~al.}(2019){Hygate}, {Kruijssen}, {Chevance}, {Schruba},
  {Haydon}, \& {Longmore}}]{hygate19}
{Hygate}, A. P.~S., {Kruijssen}, J.~M.~D., {Chevance}, M., {et~al.} 2019,
  \mnras, 488, 2800, \dodoi{10.1093/mnras/stz1779}

\bibitem[{{Kawamura} {et~al.}(2009){Kawamura}, {Mizuno}, {Minamidani},
  {Filipovi{\'c}}, {Staveley-Smith}, {Kim}, {Mizuno}, {Onishi}, {Mizuno}, \&
  {Fukui}}]{kawamura09}
{Kawamura}, A., {Mizuno}, Y., {Minamidani}, T., {et~al.} 2009, \apjs, 184, 1,
  \dodoi{10.1088/0067-0049/184/1/1}

\bibitem[{{Kennicutt}(1998)}]{kennicutt98}
{Kennicutt}, Robert~C., J. 1998, \apj, 498, 541, \dodoi{10.1086/305588}

\bibitem[{{Kennicutt} {et~al.}(2007){Kennicutt}, {Calzetti}, {Walter}, {Helou},
  {Hollenbach}, {Armus}, {Bendo}, {Dale}, {Draine}, {Engelbracht}, {Gordon},
  {Prescott}, {Regan}, {Thornley}, {Bot}, {Brinks}, {de Blok}, {de Mello},
  {Meyer}, {Moustakas}, {Murphy}, {Sheth}, \& {Smith}}]{kennicutt07}
{Kennicutt}, Robert~C., J., {Calzetti}, D., {Walter}, F., {et~al.} 2007, \apj,
  671, 333, \dodoi{10.1086/522300}

\bibitem[{{Kennicutt} \& {Evans}(2012)}]{kennicutt12}
{Kennicutt}, R.~C., \& {Evans}, N.~J. 2012, \araa, 50, 531,
  \dodoi{10.1146/annurev-astro-081811-125610}

\bibitem[{{Kim} {et~al.}(2021){Kim}, {Chevance}, {Kruijssen}, {Schruba},
  {Sandstrom}, {Barnes}, {Bigiel}, {Blanc}, {Cao}, {Dale}, {Faesi}, {Glover},
  {Grasha}, {Groves}, {Herrera}, {Klessen}, {Kreckel}, {Lee}, {Leroy}, {Pety},
  {Querejeta}, {Schinnerer}, {Sun}, {Usero}, {Ward}, \& {Williams}}]{kim21}
{Kim}, J., {Chevance}, M., {Kruijssen}, J.~M.~D., {et~al.} 2021, \mnras, 504,
  487, \dodoi{10.1093/mnras/stab878}

\bibitem[{{Kim} {et~al.}(2022){Kim}, {Chevance}, {Kruijssen}, {Leroy},
  {Schruba}, {Barnes}, {Bigiel}, {Blanc}, {Cao}, {Congiu}, {Dale}, {Faesi},
  {Glover}, {Grasha}, {Groves}, {Hughes}, {Klessen}, {Kreckel}, {McElroy},
  {Pan}, {Pety}, {Querejeta}, {Razza}, {Rosolowsky}, {Saito}, {Schinnerer},
  {Sun}, {Tomi{\v{c}}i{\'c}}, {Usero}, \& {Williams}}]{kim22}
---. 2022, \mnras, \dodoi{10.1093/mnras/stac2339}

\bibitem[{{Klessen} \& {Glover}(2016)}]{klessen16}
{Klessen}, R.~S., \& {Glover}, S. C.~O. 2016, Saas-Fee Advanced Course, 43, 85,
  \dodoi{10.1007/978-3-662-47890-5_2}

\bibitem[{{Koepferl} {et~al.}(2015){Koepferl}, {Robitaille}, {Morales}, \&
  {Johnston}}]{koepferl15}
{Koepferl}, C.~M., {Robitaille}, T.~P., {Morales}, E. F.~E., \& {Johnston},
  K.~G. 2015, \apj, 799, 53, \dodoi{10.1088/0004-637X/799/1/53}

\bibitem[{{Kruijssen} \& {Longmore}(2014)}]{kruijssen14}
{Kruijssen}, J.~M.~D., \& {Longmore}, S.~N. 2014, \mnras, 439, 3239,
  \dodoi{10.1093/mnras/stu098}

\bibitem[{{Kruijssen} {et~al.}(2019{\natexlab{a}}){Kruijssen}, {Pfeffer},
  {Crain}, \& {Bastian}}]{kruijssen19b}
{Kruijssen}, J.~M.~D., {Pfeffer}, J.~L., {Crain}, R.~A., \& {Bastian}, N.
  2019{\natexlab{a}}, \mnras, 486, 3134, \dodoi{10.1093/mnras/stz968}

\bibitem[{{Kruijssen} {et~al.}(2018){Kruijssen}, {Schruba}, {Hygate}, {Hu},
  {Haydon}, \& {Longmore}}]{kruijssen18}
{Kruijssen}, J.~M.~D., {Schruba}, A., {Hygate}, A. P.~S., {et~al.} 2018,
  \mnras, 479, 1866, \dodoi{10.1093/mnras/sty1128}

\bibitem[{{Kruijssen} {et~al.}(2019{\natexlab{b}}){Kruijssen}, {Schruba},
  {Chevance}, {Longmore}, {Hygate}, {Haydon}, {McLeod}, {Dalcanton}, {Tacconi},
  \& {van Dishoeck}}]{kruijssen19}
{Kruijssen}, J.~M.~D., {Schruba}, A., {Chevance}, M., {et~al.}
  2019{\natexlab{b}}, \nat, 569, 519, \dodoi{10.1038/s41586-019-1194-3}

\bibitem[{{Lada} \& {Lada}(2003)}]{lada03}
{Lada}, C.~J., \& {Lada}, E.~A. 2003, \araa, 41, 57,
  \dodoi{10.1146/annurev.astro.41.011802.094844}

\bibitem[{{Lee} {et~al.}(subm.)}]{LEE_PHANGSJWST}
{Lee}, J., {et~al.} subm., \apjl

\bibitem[{{Lee} {et~al.}(2022){Lee}, {Whitmore}, {Thilker}, {Deger}, {Larson},
  {Ubeda}, {Anand}, {Boquien}, {Chandar}, {Dale}, {Emsellem}, {Leroy},
  {Rosolowsky}, {Schinnerer}, {Schmidt}, {Lilly}, {Turner}, {Van Dyk}, {White},
  {Barnes}, {Belfiore}, {Bigiel}, {Blanc}, {Cao}, {Chevance}, {Congiu},
  {Egorov}, {Glover}, {Grasha}, {Groves}, {Henshaw}, {Hughes}, {Klessen},
  {Koch}, {Kreckel}, {Kruijssen}, {Liu}, {Lopez}, {Mayker}, {Meidt}, {Murphy},
  {Pan}, {Pety}, {Querejeta}, {Razza}, {Saito}, {S{\'a}nchez-Bl{\'a}zquez},
  {Santoro}, {Sardone}, {Scheuermann}, {Schruba}, {Sun}, {Usero}, {Watkins}, \&
  {Williams}}]{lee22}
{Lee}, J.~C., {Whitmore}, B.~C., {Thilker}, D.~A., {et~al.} 2022, \apjs, 258,
  10, \dodoi{10.3847/1538-4365/ac1fe5}

\bibitem[{{Leitherer} {et~al.}(1999){Leitherer}, {Schaerer}, {Goldader},
  {Delgado}, {Robert}, {Kune}, {de Mello}, {Devost}, \&
  {Heckman}}]{leitherer99}
{Leitherer}, C., {Schaerer}, D., {Goldader}, J.~D., {et~al.} 1999, \apjs, 123,
  3, \dodoi{10.1086/313233}

\bibitem[{{Leroy} {et~al.}(subm.)}]{LEROY1_PHANGSJWST}
{Leroy}, A., {et~al.} subm., \apjl

\bibitem[{{Leroy} {et~al.}(2012){Leroy}, {Bigiel}, {de Blok}, {Boissier},
  {Bolatto}, {Brinks}, {Madore}, {Munoz-Mateos}, {Murphy}, {Sandstrom},
  {Schruba}, \& {Walter}}]{leroy12}
{Leroy}, A.~K., {Bigiel}, F., {de Blok}, W.~J.~G., {et~al.} 2012, \aj, 144, 3,
  \dodoi{10.1088/0004-6256/144/1/3}

\bibitem[{{Leroy} {et~al.}(2019){Leroy}, {Sandstrom}, {Lang}, {Lewis}, {Salim},
  {Behrens}, {Chastenet}, {Chiang}, {Gallagher}, {Kessler}, \&
  {Utomo}}]{leroy19}
{Leroy}, A.~K., {Sandstrom}, K.~M., {Lang}, D., {et~al.} 2019, \apjs, 244, 24,
  \dodoi{10.3847/1538-4365/ab3925}

\bibitem[{{Leroy} {et~al.}(2021{\natexlab{a}}){Leroy}, {Schinnerer}, {Hughes},
  {Rosolowsky}, {Pety}, {Schruba}, {Usero}, {Blanc}, {Chevance}, {Emsellem},
  {Faesi}, {Herrera}, {Liu}, {Meidt}, {Querejeta}, {Saito}, {Sandstrom}, {Sun},
  {Williams}, {Anand}, {Barnes}, {Behrens}, {Belfiore}, {Benincasa},
  {Be{\v{s}}li{\'c}}, {Bigiel}, {Bolatto}, {den Brok}, {Cao}, {Chandar},
  {Chastenet}, {Chiang}, {Congiu}, {Dale}, {Deger}, {Eibensteiner}, {Egorov},
  {Garc{\'\i}a-Rodr{\'\i}guez}, {Glover}, {Grasha}, {Henshaw}, {Ho}, {Kepley},
  {Kim}, {Klessen}, {Kreckel}, {Koch}, {Kruijssen}, {Larson}, {Lee}, {Lopez},
  {Machado}, {Mayker}, {McElroy}, {Murphy}, {Ostriker}, {Pan}, {Pessa},
  {Puschnig}, {Razza}, {S{\'a}nchez-Bl{\'a}zquez}, {Santoro}, {Sardone},
  {Scheuermann}, {Sliwa}, {Sormani}, {Stuber}, {Thilker}, {Turner}, {Utomo},
  {Watkins}, \& {Whitmore}}]{leroy21_survey}
{Leroy}, A.~K., {Schinnerer}, E., {Hughes}, A., {et~al.} 2021{\natexlab{a}},
  \apjs, 257, 43, \dodoi{10.3847/1538-4365/ac17f3}

\bibitem[{{Leroy} {et~al.}(2021{\natexlab{b}}){Leroy}, {Hughes}, {Liu}, {Pety},
  {Rosolowsky}, {Saito}, {Schinnerer}, {Schruba}, {Usero}, {Faesi}, {Herrera},
  {Chevance}, {Hygate}, {Kepley}, {Koch}, {Querejeta}, {Sliwa}, {Will},
  {Wilson}, {Anand}, {Barnes}, {Belfiore}, {Be{\v{s}}li{\'c}}, {Bigiel},
  {Blanc}, {Bolatto}, {Boquien}, {Cao}, {Chandar}, {Chastenet}, {Chiang},
  {Congiu}, {Dale}, {Deger}, {den Brok}, {Eibensteiner}, {Emsellem},
  {Garc{\'\i}a-Rodr{\'\i}guez}, {Glover}, {Grasha}, {Groves}, {Henshaw},
  {Jim{\'e}nez Donaire}, {Kim}, {Klessen}, {Kreckel}, {Kruijssen}, {Larson},
  {Lee}, {Mayker}, {McElroy}, {Meidt}, {Mok}, {Pan}, {Puschnig}, {Razza},
  {S{\'a}nchez-Bl'azquez}, {Sandstrom}, {Santoro}, {Sardone}, {Scheuermann},
  {Sun}, {Thilker}, {Turner}, {Ubeda}, {Utomo}, {Watkins}, \&
  {Williams}}]{leroy21_pipe}
{Leroy}, A.~K., {Hughes}, A., {Liu}, D., {et~al.} 2021{\natexlab{b}}, \apjs,
  255, 19, \dodoi{10.3847/1538-4365/abec80}

\bibitem[{{Linden} {et~al.}(2022){Linden}, {Evans}, {Armus}, {Rich}, {Larson},
  {Lai}, {Privon}, {U}, {Inami}, {Bohn}, {Song}, {Barcos-Mu{\~n}oz},
  {Charmandaris}, {Medling}, {Stierwalt}, {Diaz-Santos}, {B{\"o}ker}, {van der
  Werf}, {Aalto}, {Appleton}, {Brown}, {Hayward}, {Howell}, {Iwasawa},
  {Kemper}, {Frayer}, {Law}, {Malkan}, {Marshall}, {Mazzarella}, {Murphy},
  {Sanders}, \& {Surace}}]{linden22}
{Linden}, S.~T., {Evans}, A.~S., {Armus}, L., {et~al.} 2022, arXiv e-prints,
  arXiv:2210.05763.
\newblock \doarXiv{2210.05763}

\bibitem[{{Lopez} {et~al.}(2014){Lopez}, {Krumholz}, {Bolatto}, {Prochaska},
  {Ramirez-Ruiz}, \& {Castro}}]{lopez14}
{Lopez}, L.~A., {Krumholz}, M.~R., {Bolatto}, A.~D., {et~al.} 2014, \apj, 795,
  121, \dodoi{10.1088/0004-637X/795/2/121}

\bibitem[{{Lu} {et~al.}(2022){Lu}, {Boyce}, {Haggard}, {Bureau}, {Liang},
  {Liu}, {Choi}, {Cappellari}, {Chemin}, {Chevance}, {Davis}, {Drissen},
  {Elford}, {Gensior}, {Kruijssen}, {Martin}, {Mass{\'e}}, {Robert}, {Ruffa},
  {Rousseau-Nepton}, {Sarzi}, {Savard}, \& {Williams}}]{lu22}
{Lu}, A., {Boyce}, H., {Haggard}, D., {et~al.} 2022, \mnras, 514, 5035,
  \dodoi{10.1093/mnras/stac1583}

\bibitem[{{Meidt} {et~al.}(2015){Meidt}, {Hughes}, {Dobbs}, {Pety}, {Thompson},
  {Garc{\'\i}a-Burillo}, {Leroy}, {Schinnerer}, {Colombo}, {Querejeta},
  {Kramer}, {Schuster}, \& {Dumas}}]{meidt15}
{Meidt}, S.~E., {Hughes}, A., {Dobbs}, C.~L., {et~al.} 2015, \apj, 806, 72,
  \dodoi{10.1088/0004-637X/806/1/72}

\bibitem[{{Meidt} {et~al.}(2018){Meidt}, {Leroy}, {Rosolowsky}, {Kruijssen},
  {Schinnerer}, {Schruba}, {Pety}, {Blanc}, {Bigiel}, {Chevance}, {Hughes},
  {Querejeta}, \& {Usero}}]{meidt18}
{Meidt}, S.~E., {Leroy}, A.~K., {Rosolowsky}, E., {et~al.} 2018, \apj, 854,
  100, \dodoi{10.3847/1538-4357/aaa290}

\bibitem[{{Miura} {et~al.}(2012){Miura}, {Kohno}, {Tosaki}, {Espada}, {Hwang},
  {Kuno}, {Okumura}, {Hirota}, {Muraoka}, {Onodera}, {Minamidani}, {Komugi},
  {Nakanishi}, {Sawada}, {Kaneko}, \& {Kawabe}}]{miura12}
{Miura}, R.~E., {Kohno}, K., {Tosaki}, T., {et~al.} 2012, \apj, 761, 37,
  \dodoi{10.1088/0004-637X/761/1/37}

\bibitem[{{Onodera} {et~al.}(2010){Onodera}, {Kuno}, {Tosaki}, {Kohno},
  {Nakanishi}, {Sawada}, {Muraoka}, {Komugi}, {Miura}, {Kaneko}, {Hirota}, \&
  {Kawabe}}]{onodera10}
{Onodera}, S., {Kuno}, N., {Tosaki}, T., {et~al.} 2010, \apjl, 722, L127,
  \dodoi{10.1088/2041-8205/722/2/L127}

\bibitem[{{Padoan} {et~al.}(2017){Padoan}, {Haugb{\o}lle}, {Nordlund}, \&
  {Frimann}}]{padoan17}
{Padoan}, P., {Haugb{\o}lle}, T., {Nordlund}, {\r{A}}., \& {Frimann}, S. 2017,
  \apj, 840, 48, \dodoi{10.3847/1538-4357/aa6afa}

\bibitem[{{Pan} {et~al.}(2022){Pan}, {Schinnerer}, {Hughes}, {Leroy}, {Groves},
  {Barnes}, {Belfiore}, {Bigiel}, {Blanc}, {Cao}, {Chevance}, {Congiu}, {Dale},
  {Eibensteiner}, {Emsellem}, {Faesi}, {Glover}, {Grasha}, {Herrera}, {Ho},
  {Klessen}, {Kruijssen}, {Lang}, {Liu}, {McElroy}, {Meidt}, {Murphy}, {Pety},
  {Querejeta}, {Razza}, {Rosolowsky}, {Saito}, {Santoro}, {Schruba}, {Sun},
  {Tomicic}, {Usero}, {Utomo}, \& {Williams}}]{pan22}
{Pan}, H.-A., {Schinnerer}, E., {Hughes}, A., {et~al.} 2022, arXiv e-prints,
  arXiv:2201.01403.
\newblock \doarXiv{2201.01403}

\bibitem[{{Rodriguez} {et~al.}(subm.)}]{RODRIGUEZ_PHANGSJWST}
{Rodriguez}, J., {et~al.} subm., \apjl

\bibitem[{{Rosolowsky} {et~al.}(2021){Rosolowsky}, {Hughes}, {Leroy}, {Sun},
  {Querejeta}, {Schruba}, {Usero}, {Herrera}, {Liu}, {Pety}, {Saito},
  {Be{\v{s}}li{\'c}}, {Bigiel}, {Blanc}, {Chevance}, {Dale}, {Deger}, {Faesi},
  {Glover}, {Henshaw}, {Klessen}, {Kruijssen}, {Larson}, {Lee}, {Meidt}, {Mok},
  {Schinnerer}, {Thilker}, \& {Williams}}]{rosolowsky21}
{Rosolowsky}, E., {Hughes}, A., {Leroy}, A.~K., {et~al.} 2021, \mnras, 502,
  1218, \dodoi{10.1093/mnras/stab085}

\bibitem[{{Schinnerer} {et~al.}(2019){Schinnerer}, {Hughes}, {Leroy}, {Groves},
  {Blanc}, {Kreckel}, {Bigiel}, {Chevance}, {Dale}, {Emsellem}, {Faesi},
  {Glover}, {Grasha}, {Henshaw}, {Hygate}, {Kruijssen}, {Meidt}, {Pety},
  {Querejeta}, {Rosolowsky}, {Saito}, {Schruba}, {Sun}, \&
  {Utomo}}]{schinnerer19}
{Schinnerer}, E., {Hughes}, A., {Leroy}, A., {et~al.} 2019, \apj, 887, 49,
  \dodoi{10.3847/1538-4357/ab50c2}

\bibitem[{{Schruba} {et~al.}(2010){Schruba}, {Leroy}, {Walter}, {Sandstrom}, \&
  {Rosolowsky}}]{schruba10}
{Schruba}, A., {Leroy}, A.~K., {Walter}, F., {Sandstrom}, K., \& {Rosolowsky},
  E. 2010, \apj, 722, 1699, \dodoi{10.1088/0004-637X/722/2/1699}

\bibitem[{{Thilker} {et~al.}(subm.)}]{THILKER_PHANGSJWST}
{Thilker}, D., {et~al.} subm., \apjl

\bibitem[{{Turner} {et~al.}(2022){Turner}, {Dale}, {Lilly}, {Boquien}, {Deger},
  {Lee}, {Whitmore}, {Anand}, {Benincasa}, {Bigiel}, {Blanc}, {Chevance},
  {Emsellem}, {Faesi}, {Glover}, {Grasha}, {Hughes}, {Klessen}, {Kreckel},
  {Kruijssen}, {Leroy}, {Pan}, {Rosolowsky}, {Schruba}, \&
  {Williams}}]{turner22}
{Turner}, J.~A., {Dale}, D.~A., {Lilly}, J., {et~al.} 2022, \mnras, 516, 4612,
  \dodoi{10.1093/mnras/stac2559}

\bibitem[{{Verley} {et~al.}(2009){Verley}, {Corbelli}, {Giovanardi}, \&
  {Hunt}}]{verley09}
{Verley}, S., {Corbelli}, E., {Giovanardi}, C., \& {Hunt}, L.~K. 2009, \aap,
  493, 453, \dodoi{10.1051/0004-6361:200810566}

\bibitem[{{Viaene} {et~al.}(2017){Viaene}, {Baes}, {Tamm}, {Tempel}, {Bendo},
  {Blommaert}, {Boquien}, {Boselli}, {Camps}, {Cooray}, {De Looze}, {De Vis},
  {Fern{\'a}ndez-Ontiveros}, {Fritz}, {Galametz}, {Gentile}, {Madden}, {Smith},
  {Spinoglio}, \& {Verstocken}}]{viaene17}
{Viaene}, S., {Baes}, M., {Tamm}, A., {et~al.} 2017, \aap, 599, A64,
  \dodoi{10.1051/0004-6361/201629251}

\bibitem[{{Virtanen} {et~al.}(2020){Virtanen}, {Gommers}, {Oliphant},
  {Haberland}, {Reddy}, {Cournapeau}, {Burovski}, {Peterson}, {Weckesser},
  {Bright}, {van der Walt}, {Brett}, {Wilson}, {Millman}, {Mayorov}, {Nelson},
  {Jones}, {Kern}, {Larson}, {Carey}, {Polat}, {Feng}, {Moore}, {VanderPlas},
  {Laxalde}, {Perktold}, {Cimrman}, {Henriksen}, {Quintero}, {Harris},
  {Archibald}, {Ribeiro}, {Pedregosa}, {van Mulbregt}, \& {SciPy 1. 0
  Contributors}}]{virtanen20}
{Virtanen}, P., {Gommers}, R., {Oliphant}, T.~E., {et~al.} 2020, Nature
  Methods, 17, 261, \dodoi{10.1038/s41592-019-0686-2}

\bibitem[{{Ward} {et~al.}(2020){Ward}, {Chevance}, {Kruijssen}, {Hygate},
  {Schruba}, \& {Longmore}}]{ward20_HI}
{Ward}, J.~L., {Chevance}, M., {Kruijssen}, J.~M.~D., {et~al.} 2020, \mnras,
  497, 2286, \dodoi{10.1093/mnras/staa1977}

\bibitem[{{Ward} {et~al.}(2022){Ward}, {Kruijssen}, {Chevance}, {Kim}, \&
  {Longmore}}]{ward22}
{Ward}, J.~L., {Kruijssen}, J.~M.~D., {Chevance}, M., {Kim}, J., \& {Longmore},
  S.~N. 2022, arXiv e-prints, arXiv:2209.05541.
\newblock \doarXiv{2209.05541}

\bibitem[{{Whitmore} {et~al.}(subm.)}]{WHITMORE_PHANGSJWST}
{Whitmore}, B., {et~al.} subm., \apjl

\bibitem[{{Whitmore} {et~al.}(2014){Whitmore}, {Brogan}, {Chandar}, {Evans},
  {Hibbard}, {Johnson}, {Leroy}, {Privon}, {Remijan}, \& {Sheth}}]{whitmore14}
{Whitmore}, B.~C., {Brogan}, C., {Chandar}, R., {et~al.} 2014, \apj, 795, 156,
  \dodoi{10.1088/0004-637X/795/2/156}

\bibitem[{{Williams} {et~al.}(1994){Williams}, {de Geus}, \&
  {Blitz}}]{williams94}
{Williams}, J.~P., {de Geus}, E.~J., \& {Blitz}, L. 1994, \apj, 428, 693,
  \dodoi{10.1086/174279}

\bibitem[{{Williams} {et~al.}(2019){Williams}, {Baes}, {De Looze},
  {Rela{\~n}o}, {Smith}, {Verstocken}, \& {Viaene}}]{williams19}
{Williams}, T.~G., {Baes}, M., {De Looze}, I., {et~al.} 2019, \mnras, 487,
  2753, \dodoi{10.1093/mnras/stz1441}

\bibitem[{{Zabel} {et~al.}(2020){Zabel}, {Davis}, {Sarzi}, {Nedelchev},
  {Chevance}, {Kruijssen}, {Iodice}, {Baes}, {Bendo}, {Corsini}, {De Looze},
  {de Zeeuw}, {Gadotti}, {Grossi}, {Peletier}, {Pinna}, {Serra}, {van de
  Voort}, {Venhola}, {Viaene}, \& {Vlahakis}}]{zabel20}
{Zabel}, N., {Davis}, T.~A., {Sarzi}, M., {et~al.} 2020, \mnras, 496, 2155,
  \dodoi{10.1093/mnras/staa1513}

\end{thebibliography}

\suppressAffiliationsfalse
\allauthors

\end{document}